\documentclass[graybox,natbib,nosecnum]{svmult}
\bibpunct{(}{)}{;}{a}{}{,} 

\pdfoutput=1   

\usepackage{amsmath}

\usepackage{mathptmx}       
\usepackage{helvet}         
\usepackage{courier}        
\usepackage{type1cm}        
\usepackage{subfig}
\usepackage{makeidx}         
\usepackage{graphicx}        
\usepackage{multicol}        
\usepackage[bottom]{footmisc}
\usepackage[normalem]{ulem}	
\usepackage{floatrow}
\usepackage{soul}   

\usepackage{txfonts}

\usepackage{IEEEtrantools}

\usepackage{wasysym}
\usepackage{marvosym}

\usepackage{mathtools}

\usepackage{hyperref}             
\hypersetup{pdfauthor=Christoph Mordasini}
\hypersetup{backref=true, pagebackref=true, hyperindex=true, breaklinks=true,colorlinks=true,urlcolor=blue, linkcolor=blue,  citecolor=blue,pagecolor=red, bookmarks=true, bookmarksopen=true}

\def\mearth{M_\oplus}
\def\rearth{R_\oplus}
\def\msun{M_\odot}

\def\mcore{M_{\rm core}}

\def\f1{f_{\rm I}}
\def\mstar{M_*}

\def\menv{M_{\rm env}}

\def\beq{\begin{equation}}
\def\eeq{\end{equation}}

\def\t2{\tau_{\rm II}}

\def\sigmas0{\Sigma_{\rm s,0}}
\def\mj{M_{\textrm{\tiny \jupiter }}}

\newcommand{\rj}{R_{\textrm{\tiny \jupiter}}}

\def\tKH{\tau_{\rm KH}}

\makeindex             

\begin{document}

\title*{Planetary population synthesis}
\author{Christoph Mordasini}
\institute{Christoph Mordasini \at Physikalisches Institut, University of Bern, Gesellschaftsstrasse~6
CH 3012 Bern, Switzerland. \email{christoph.mordasini@space.unibe.ch}}
%
%
\maketitle

\abstract{In  {stellar} astrophysics, the technique of   population synthesis has been successfully used for several decades. For  planets,  {it} is in contrast  {still} a young method which only became important in recent years because of the rapid increase of the number of known extrasolar planets, and the associated growth of statistical observational constraints. With planetary population synthesis, the theory of planet formation and evolution can be put to the test against these  constraints. In this review  {of}  planetary population synthesis{, w}e first briefly list key observational constraints. Then, the work flow in the method and its two main components are presented, namely  global end-to-end models {that predict planetary system properties directly from protoplanetary disk properties} and probability distributions for the{se} initial conditions. An overview of various population synthesis models in the literature is given. The sub-models for the physical processes considered in global models are described: the  evolution of the protoplanetary disk, the planets' accretion of solids and gas, orbital migration, and N-body interactions among  concurrently growing protoplanets. {Next, typical population synthesis} results are illustrated {in the form of new} syntheses obtained with the latest generation of the Bern model. Planetary formation tracks, the distribution of planets in the mass-distance and radius-distance plane, the planetary mass function, and the distributions of planetary radii, semimajor axes, and luminosities are shown{, linked to underlying physical processes,}  and compared with their observational counterparts. {We finish by highlighting the most important predictions made by population synthesis models and discuss the lessons learned from these predictions - both those later observationally confirmed and those rejected. }}

\section{Confronting theory and observation}\label{sect:confrontingtheoryobs}
Since the discovery of the first extrasolar planet around a solar-like star by \citet{mayorqueloz1995}, it has become clear from observations that the population of extrasolar planets is characterized by  extreme diversity. This diversity in terms of planetary masses, orbital distances, system architectures, internal compositions etc. was not  anticipated by earlier theoretical models of planet formation \citep[e.g.,][]{boss1995} that were based on just one planetary system, our own solar system, indicating important shortcomings in the theory.

Since 1995, the number of known extrasolar planets has increased rapidly, reaching now several thousand \citep{schneiderdedieu2011,wrightfakhouri2011}. This allows {one} to study the extrasolar planets as a statistical population instead of single objects only, even though the study of a benchmark individual planetary systems (including the solar system) continues to be key to understand planet formation as well. This planetary population is characterized by a number of statistical distribution (e.g., of the mass or eccentricity), dependencies on host star properties (like the stellar metallicity), and correlations between these quantities. These statistical constraints provide a rich  data set with which the theoretical predictions of population synthesis models can be confronted.

The basic idea behind the planetary population synthesis method is that the observed diversity of extrasolar planets is due to a diversity in the initial conditions, the protoplanetary disks \citep[e.g.,][]{andrewswilner2010}. While it is typically difficult to observe the process of planet formation directly (except for a handful special cases, e.g., \citealt{sallumfollette2015}), in a numerical model the link between disk and planetary system properties can be established with so-called global model. This class of models directly predicts the final (potentially observable) properties of synthetic extrasolar planets based on the properties of their parent synthetic protoplanetary disk. For this, such  global end-to-end models build on simplified results of many different detailed model{s} for individual physical processes of planet formation, like accretion and migration. In global models, these individual processes are linked together, which is a source of considerable complexity, even for relatively simple sub-models. Thanks to this approach, the population-wide, statistical consequences of an individual physical description (like orbital migration, e.g., \citealt{massetcasoli2010}) become clear and can be statistically compared with the observed population (e.g., the semimajor axis distribution or the frequency of mean motion resonances). This means that first, theoretical models of a specific process can be put to the observational test which is otherwise often difficult as we can only observe the combined effect of all acting processes, and second that the full wealth of observational data (the entire statistical information coming from different observational techniques like radial velocities, transits, direct imaging, microlensing, ...) can be used to constrain theoretical planet formation models. This also avoids that models are constructed that can only describe specific types of systems, but fail for many others, as illustrated for the case of the solar system mentioned above.

As global models in population syntheses are in the end nothing else than coupled  agglomerates of other specialized models, it is clear that the predictions of population syntheses directly reflect the state of the field of planet formation theory as a whole, which is exactly their purpose. This means that as our understanding of planet formation changes, so do the population synthesis models. 

In {planet} formation, fundamental physical {processes} governing planet formation are currently still uncertain. An important example for this are the processes that drive accretion in protoplanetary disk  (classical MRI-driven viscous accretion, MHD winds, e.g., \citealt[][]{bai2016}), which has via different disk structures strong consequences for planet formation  \citep{ogiharamorbidelli2015}. Another important example is the relative importance of solid building blocks of various sizes ranging from cm-sized pebbles to classical 100 km-sized planetesimals \citep[][]{ormel2017}. It could appear that given such large uncertainties, currently no meaningful global models can therefore be constructed - but actually, the argument must be turned around: it is  with population syntheses that these different theories can be put to the observational test to identify which ones lead to synthetic populations that agree or disagree with observations, and to improve in this way the understanding of how planets form, which is the final goal. 

This chapter is organized as follows: we first discuss the most important observational constraints, then describe the method of population synthesis including a short overview of the input physics currently considered in global models. We then turn to the discussion of the most important results including the comparison with observations. We conclude the chapter with the discussion of tests of specific sub-models and predictions for future instruments and surveys.

For further information on the method, the reader may consider the  reviews of \citet{benzida2013} and \citet{mordasinimolliere2015}. Two other relevant publications for the (initial) development of the method are  \citet{idalin2004a} and \citet{mordasinialibert2009a}.

\section{Statistical observational constraints}\label{sect:statobsconst}
The {number} and type of  observational statistical constraints available for comparisons is in principle very large and multifaceted \citep[for reviews, see][]{udrysantos2007,winnfabrycky2015}. However, there are a number of key constraints the comparison to which population syntheses have  traditionally focussed on. 

These key constraints are usually the results of large observational surveys, both from the ground and  space. Important surveys and publications analyzing them are, e.g., the HARPS high precision radial velocity survey \citep{mayormarmier2011}, the  Keck \& Lick radial velocity survey \citep{howardmarcy2010}, the CoRoT \citep{moutoudeleuil2013} and Kepler transit surveys \citep{coughlinmullally2016}, the various direct imaging \citep[][]{bowler2016}, or the microlensing surveys \citep[][]{cassankubas2012}. The high importance of surveys stems from the fact that they have a well known observational bias. This makes it possible to correct for it and to infer the underlying actual distributions that are predicted by the theoretical models.  

All these different techniques put constraints on different aspects of the global models. Especially when they are combined, they are highly constraining even for global models that often have a significant number of free parameters as the combined data carries so much constraining information. 

The constraints can be grouped into three classes: the frequency of different planet types, the distribution functions of planetary properties, and correlations with stellar properties. We next give a short overview of these observational constraints.

\subsection{Frequencies of planet types}\label{sect:frequenciesobs}
\begin{itemize}
\item The frequency of hot Jupiters around solar-like stars is about 0.5-1\% \citep{howardmarcy2010,mayormarmier2011}. 
\item The frequency of giant planets within 5-10 AU is 10-20\% for FGK stars \citep{cummingbutler2008,mayormarmier2011}. The giant planets have a multiplicity rate of about 50\% \citep{bryanknutson2016}.
\item There is a high frequency (20-50\%) of close-in (fractions of an AU) low-mass (a few Earth masses) respectively small ($R\lesssim4\rearth$) super-Earth and sub-Neptunian planets from high-precision radial velocity \citep{mayormarmier2011} and the Kepler survey \citep{fressintorres2013a,petigurahoward2013}. These planets are often found in tightly packed multiple systems. Planetary systems clearly different from the solar system are thus very frequent.
\item There is a low frequency on the 1\% level of detectable (i.e., sufficiently luminous) massive giant planets at distances of tens to hundreds of AU. This means that the frequency of giant planets must somewhere drop with orbital distance by about a factor ten. The occurrence rate is  likely positively correlated with the stellar mass \cite{bowler2016}.
\item There is a high frequency of cold, roughly Neptunian-mass planets around M dwarfs as found by microlensing surveys \citep[][]{cassankubas2012}.
\item There is a very high total fraction of stars with detectable planets of $\sim$75\% as indicated by high-precision radial velocity searches with a $\sim$ 1 m/s precision \citep{mayormarmier2011}. At least in the solar neighborhood, stars with planets are thus the rule.
\end{itemize} 

\begin{figure}[h]
    \centering
    \includegraphics[width=\textwidth]{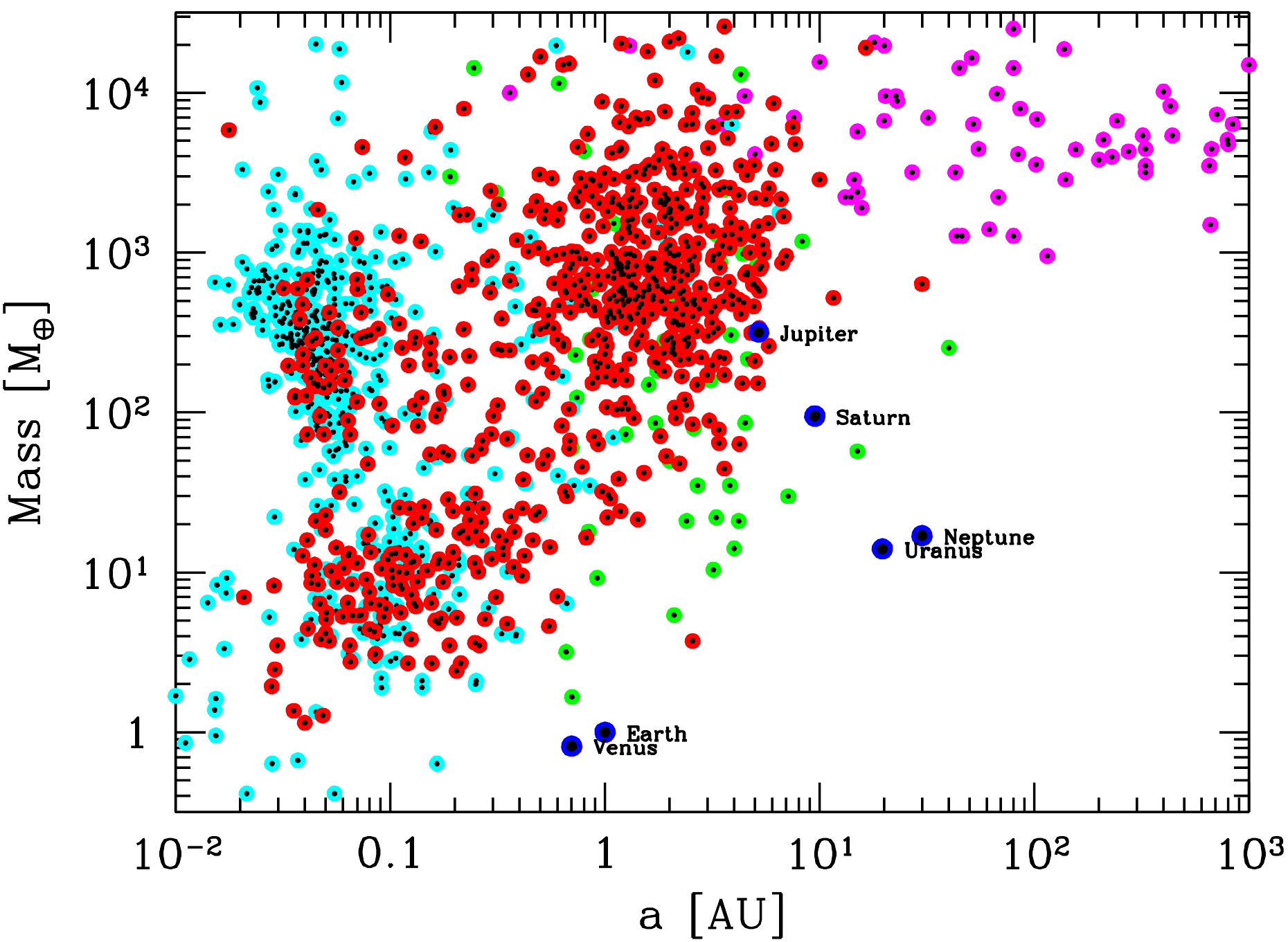}
   \caption{Mass-distance diagram of confirmed planets in ``The Extrasolar Planets Encyclopedia'' \citep{schneiderdedieu2011} as of 2017. Red, cyan, magenta, and green points indicate planets detected by the radial velocity, transits, direct imaging, and microlensing technique, respectively. The planets of the solar system are also shown for comparison.    }\label{CMaMEpoch2017}
\end{figure}

\subsection{Distributions of planetary properties }\label{sect:distribobsprop}
\begin{itemize}
\item One of the most important diagrams is the observed two-dimensional distribution of planets in the mass-distance (or radius-distance) plane, revealing a number of pile-ups and deserts  (see Fig. \ref{CMaMEpoch2017}). For the comparison with the synthetic populations, it is of paramount importance to keep in mind that the observed diagram gives a highly distorted impression of the actual population because of the detection biases of the different techniques. Hot Jupiters, for example, appear to be frequent in this plot. But the plot still illustrates the enormous diversity in the outcome of the planet formation process. At the same time,  it also indicates that  there is some structure. 
\item The mass function is approximately flat in log space in the giant planet regime \citep{marcybutler2005} for masses between 30 $\mearth$ and about 4 $\mj$ {(where 1 $\mj$ is the mass of Jupiter)}. At even higher masses, there is a drop in frequency \citep{santosadibekyan2017}. The upper end of the planetary mass function is poorly known, but might lie around 30 $\mj$ \citep{sahlmannsegransan2011}. Towards the lower masses, at around 30 $\mearth$, there is a break in the mass function and a strong increase of the frequency towards smaller masses \citep{mayormarmier2011}. The mass function below a few Earth masses is currently unknown. 
\item The semimajor axis distribution of giant planets consists of a local maximum at a period around 4 days caused by the hot Jupiters, a less populated region further out (the period valley) and finally an upturn at around 1 AU \citep{udrymayor2003}. The frequency seems to be decreasing beyond 3-10 AU \citep{bryanknutson2016}.
\item The eccentricity distribution is, in contrast to the solar system with its very low eccentricities, broad, including some planets with eccentricities that exceed 0.9. The upper part of the distribution follows approximately a Rayleigh distribution, as expected from  gravitational planet-planet interactions \citep{jurictremaine2008}{, indicating together with several other points that in some systems strong dynamical interactions occurred (see the discussion in \citealt{winnfabrycky2015})}.  A significant fraction of orbits are however also consistent with being circular. Eccentricities of lower mass planets ($\lesssim30\mearth$) are usually restricted to lower values $\leq 0.5$ \citep{mayormarmier2011}.
\item The radius distribution of confirmed (Kepler) planets has a local maximum at around 1 Jovian radius as expected from the theoretical mass-radius relation \citep{mordasinialibert2012c}, followed by a distribution that is approximately flat in $\log(R)$ at intermediate radii of 4-10 $\rearth$. Below this radius, there is strong increase in frequency \citep{fressintorres2013a,petigurahoward2013}. At about 1.7 $\rearth$, there is a local minimum in the radius histogram \citep{fultonpetigura2017} separating super-Earths from sub-Neptunes. This could be due to atmospheric escape of primordial H/He envelopes \citep[][]{owenwu2017,jinmordasini2018}.
\end{itemize}

\subsection{Correlations with stellar properties}\label{sect:obscorrstellarprops}
\begin{itemize}
\item The best known correlation of planetary and stellar properties is the increase of the frequency of giant planets with host star metallicity \citep[][]{gonzalez1997,santosisraelian2004,fischervalenti2005}. In the super-solar metallicity domain, the frequency of giant planet increases approximately by a factor ten when going from [Fe/H]=0 to [Fe/H]=0.5. This is often taken as indication that core accretion is the dominant mode of giant planet formation \citep{idalin2004,mordasinialibert2012a}. The frequency of low-mass planets is in contrast independent of metallicity  \citep{mayormarmier2011}.
\item The frequency of giant planets is lower for lower mass stars and around 2\% for M-dwarfs \citep{bonfilsdelfosse2013}. For stellar masses higher than 1 $\msun$, the frequency first increases to reach a maximum at around 2 $\msun$, followed by a  rapid drop for $\mstar \gtrsim 2.7 \msun$ \citep{reffertbergmann2015}. 
\item Statistical correlations with stellar age are  not yet well explored, but a number of detections of close-in planets around T-Tauri and young PMS  stars have occurred \citep{mannnewton2016,davidhillenbrand2016,donatimoutou2016,yudonati2017}. They show that close-in massive planets already exist after a few Myr, likely indicating  orbital migration via planet-disk interactions.  Hot Jupiters might be more frequent around T Tauri stars than main sequence stars \citep{yudonati2017}. At large orbital distances, direct imaging also probes young planets with ages of a few 10 Myr. The PLATO survey will put statistical constraints on the temporal evolution of the population of transiting planets, adding a new temporal dimension to the constraints. 
\end{itemize}

\section{Population synthesis method}\label{sect:popsynthmethod}
In this section, we review the general workflow in the method, the past development of population syntheses models, the physical processes considered in global formation and evolution models, and finally the probability distributions of the initial conditions.

\subsection{Workflow of the population synthesis method}\label{sect:workflow}
\begin{figure}[h]
    \centering
    \includegraphics[width=\textwidth]{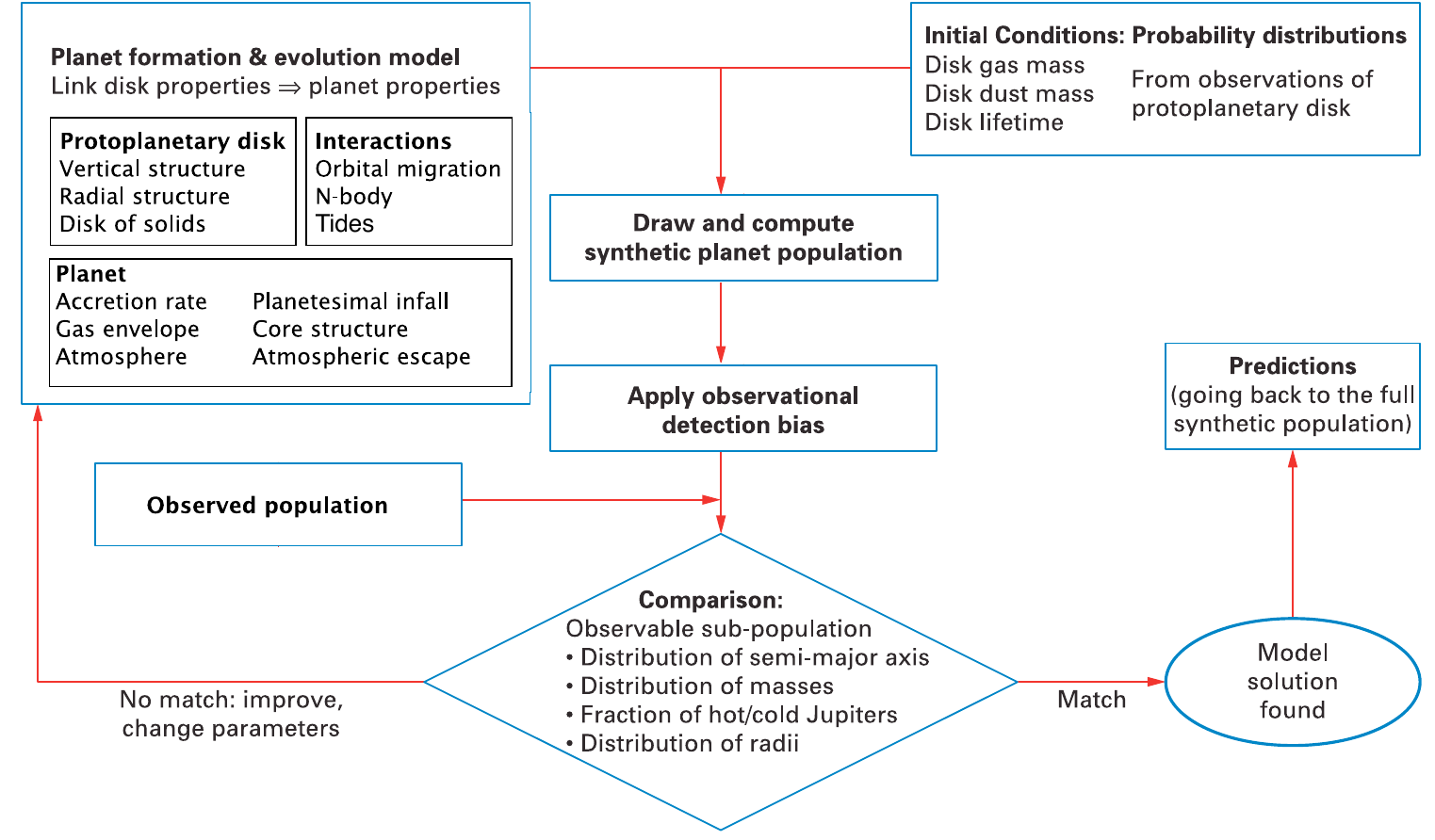}
   \caption{Elements and work flow of a planetary population synthesis framework \citep[updated from][]{mordasinimolliere2015}.}\label{CMpopsynthworkflow}
\end{figure}

The general workflow of the planetary population synthesis method is shown in Fig. \ref{CMpopsynthworkflow}. There are three main elements: first, and most importantly, the global model that predicts planetary properties directly based on disk properties. Second, the Monte Carlo distributions for the initial conditions of the global models that are derived from disk observations, from reconstructions of the disk properties in an equivalent way as done for the minimum mass solar nebula, or from theoretical arguments. Third, tools to apply the observational detection bias and to conduct the statistical comparison with the observed population.  In general, this comparison will reveal differences between the theory and observations, which are then tracked back to assumptions about the governing physical processes as implemented in the model as well as the setting of model parameters. In case that the synthetic population matches the observations at least regarding a certain aspect, the synthetic population can also be used to make predictions about aspects that cannot yet be observed, including the expected yield of future surveys.

\subsection{Overview of population synthesis models in the literature}\label{sect:overviewpopmodels}
In other fields of astrophysics, (stellar) population synthesis is a well-established technique  for several decades \citep[e.g.,][]{bruzualcharlot2003}, while for planets, it is still a {recent} approach. The construction of planetary population synthesis models was triggered by the rapidly increasing number of known extrasolar planets. In this section, we review  past and present developments of such models by various groups. Early models were all based on the classical core accretion paradigm where the solids are accreted in the form of planetesimals \citep{perricameron1974,mizuno1980,bodenheimerpollack1986,pollackhubickyj1996}. More recently, models were also based on core accretion with pebbles \citep{ormelklahr2010,johansenlambrechts2017}, and on planet formation via gravitational instability \citep{kuiper1951,cameron1978,boss1997}.

\begin{enumerate}
\item The Ida \& Lin models. The pioneering population synthesis calculation of \citet{idalin2004a} contained for the first time all the basic elements of population synthesis shown in Fig. \ref{CMpopsynthworkflow}, namely a purpose-built - and therefore fast - global planet formation model based on the  core accretion paradigm, and a variation of  the initial conditions in a Monte Carlo way. The effects of planetesimal accretion, parameterized gas accretion, and Type II orbital migration in simple power-law disks were considered. As most first-generation population synthesis models, the one-embryo-per-disk approximation was used. Later works added Type I migration \citep{idalin2008c}, a density enhancement due to a dead zone at the iceline \citep{idalin2008}, and finally a semi-analytical statistical treatment of the dynamical interactions of several concurrently growing protoplanets \citep{idalin2010,idalin2013}. 
 
\item The Bern Model. Building on the Alibert, Mordasini \& Benz (\citeyear{alibertmordasini2004,alibertmordasini2005}) model for giant planet formation in the solar system, \citet{mordasinialibert2009a,mordasinialibert2009b} presented population syntheses that included quantitative statistical comparisons with observations. Compared to the Ida \& Lin models, the Bern model explicitly solves the (partial) differential for the structure and evolution of the protoplanetary disk and the planets' interior structure, rather then using  power-law solutions. This  has the implication of substantially higher computational costs.  Subsequent improvements addressed the  structure of the protoplanetary disk \citep{fouchetalibert2012}, the solid accretion rate \citep{fortieralibert2013}, and the type I migration description \citep{dittkristmordasini2014}. The model was extended to include the planets' post-formation thermodynamic evolution (cooling and  contraction) over Gyrs timescales \citep{mordasinialibert2012b}, as well as atmospheric escape \citep{jinmordasini2014}. This makes it possible to predict directly also radii and luminosities instead of masses only. Also these models originally used the one-embryo-per-disk approximation. The concurrent formation of multiple protoplanets interacting via an explicit N-body integrator was added in \citet{alibertcarron2013}.

\item The models of \citet{hasegawapudritz2011a,hasegawapudritz2012,hasegawapudritz2013}  combine a planet formation model based {initially} on the Ida\& Lin models with power-law disks with inhomogeneities or the analytical disk model of \citet{chambers2009} {and \citet{cridlandpudritz2016}}. These models emphasize the importance of ``planet traps'', i.e., special locations in the disk where orbital migration is slowed down or stopped due to transitions in the disk. These transitions are the edge of the MRI-dead zone, icelines, and the transition from the viscously heated to the irradiation-dominated region in the disk. Later updates \citep{alessipudritz2017,cridlandpudritz2017} include models for the dust physics, astrochemistry, and radiative transfer.

\item While not used in population syntheses but in parameter studies, the models of \citet{hellarynelson2012,colemannelson2014,colemannelson2016}  are global models that combine an N-body integrator with a 1D  model for the disk's structure and evolution and the planets' orbital migration. In contrast to other models, the planetesimals are directly included in the N-body as massless test particles, and not simply represented as a surface density. Early models use fits to the results of \citet{movshovitzbodenheimer2010} for the planets' gas accretion rate while later  models \citep{colemanpapaloizou2017} calculate it by solving 1D structure equations. Similar global models were also presented by \citet{thommesmatsumura2008a}.

\item Based on the global model of \citet{bitschLambrechts2015}, \citet{ndugubitsch2018} presented population syntheses based on the core accretion paradigm where the cores grow by the accretion of pebbles instead of planetesimals. \citet{ndugubitsch2018} focussed on the effect of the stellar  cluster environment. The gas disk structure is obtained from 2D simulations including viscous heating and stellar irradiation assuming a radially constant mass flux \citep{bitschjohansen2015}. The planets' gas accretion rate is given by analytical results of \citet{pisoyoudin2014}.  The cores grow by the accretion of mm-cm sized drifting pebbles \citep{lambrechtsjohansen2012,lambrechtsjohansen2014}. The model uses  the one-embryo-per-disk approach, such that N-body interactions are neglected, while type I and II migration are included.

\item An increasing number of population synthesis calculations are also based on variants of the gravitational instability model for giant planet formation \citep[e.g.,][]{forganrice2013a,nayakshinfletcher2015,muellerhelled2018}. Similar to the core accretion models, these global models couple simple semi-analytic sub-models of disc evolution, disk fragmentation, initial embryo mass, gas accretion and loss (for example by tidal downsizing, \citealt{Nayakshin2010a}), orbital migration, grain growth, formation of solid cores by sedimentation, and recently, the N-body interaction of several fragments \citep{forganhall2018}.
\end{enumerate}

In the remainder of the article we concentrate on the population synthesis models based on the core accretion paradigm.

\subsection{Global models: simplified but  linked}\label{sect:globalmodels}
The core accretion paradigm states that giant planets form in a two-step process. First a so-called critical core is built (with a mass of about 10 $\mearth$), which then triggers the accretion of the gaseous envelope. This happens in evolving disks of gas and solids in which also other protoplanets grow, leading to dynamical interactions. The gas disk and the protoplanets exchange angular moment, so that orbital migration occurs. All these processes occur on similar timescales, meaning that they need to be considered in a self-consistently coupled fashion. A global planet formation model must thus consider this minimal set of physical processes \citep{benzida2013}:
\begin{enumerate}
\item The structure and evolution of the protoplanetary gas disk 
\item The structure and evolution of the disk of solids (dust, pebbles, planetesimals)
\item The accretion of solids leading to the growth of the planetary solid core 
\item The accretion of H/He leading to the growth of the planetary gaseous envelope 
\item Orbital migration resulting from the exchange of angular momentum 
\item N-body interaction among (proto)planets
\end{enumerate}
Additional sub-models may describe the internal structure of the core and envelope, the structure of the planetary atmosphere, the interaction of infalling planetesimals and pebbles with the protoplanets' envelope, the evolution of the star,  or the loss of the gaseous envelope during the evolutionary phase, for example via atmospheric escape \citep[e.g.,][]{jinmordasini2014}. 

\subsection{Low-dimensional approximation}\label{sect:lowdimapprox}
{For a statistical approach like population synthesis where hundreds of planetary systems must be simulated over timescales of many millions of years during the formation epoch, and even for billions of years during the evolution phase, it is currently not possible to use detailed multi-dimensional  hydrodynamic simulations possibly even including radiative transfer and magnetic fields because of computational time limitations.}

{Instead, the sub-models are either parameterized based on the results of  detailed models or solve differential equations describing low-dimensional approximations like 1D spherically symmetric hydrostatic planet interior equations or 1D axisymmetric protoplanetary disk evolution equations. A key challenge of population synthesis is thus to ``distill'' the insights  from 2D or 3D detailed  models of one specific process (like orbital migration or pebble accretion) into simpler computationally efficient approximations, that however still correctly capture the essence of the governing physical mechanism. Exploring which approximations are possible without losing the essence is an ongoing challenge for population synthesis models.}

{On the other hand, the fact that the different processes are considered in a self-consistent coupled fashion over a long timescale is a strong aspect of global models, as it captures the non-linear interactions of the different processes as they are occurring also in nature. This coupling is a source of considerable complexity of the models, even for relatively simple individual sub-models.}

We next briefly discuss some of these sub-models. More detailed descriptions can be found in \citet{benzida2013} and \citet{mordasinimarleau2017}. 

\subsubsection{Disk models}
The disk model describes the evolution of the surface density of gas and solids. It also gives the gas temperature, pressure, and vertical scale height. These quantities and their radial derivatives enter into the other sub-models in multiple ways, making the disk model a key component of a global model. The gas disk properties for example determine as outer boundary conditions the gas accretion rate of protoplanets, enter into the migration rates, control the aerodynamic behavior of small particles, or the damping of the random velocities of the planetesimals.

The simplest way of setting up a parameterized (gas) disk model is a power law approach inspired by the {minimum mass solar nebula} MMSN \citep{weidenschilling1977,hayashi1981}, as used in the original models of  \citet{idalin2004a}. {In the MMSN approach, the present-day positions, masses, and compositions of the solar system planets are used to reconstruct the radial distribution of matter in the solar nebula, assuming in situ growth.} In such models, the (initial) surface density of gas $\Sigma$ as a function of distance from the star $r$ is given as
\beq\label{eqSigmapw}
\Sigma(r)=\Sigma_{0} \left(\frac{r}{\rm 1 AU}\right)^{-3/2}.
\eeq
In a population synthesis, the normalization constant $\Sigma_{\rm 0}$ is varied as a Monte Carlo variable to represent disk of different masses {(see Sect. \nameref{sect:probdists})} with $\Sigma_{\rm0}\approx2400$ g/cm$^{2}$ corresponding for example to the surface density in the MMSN. In such simple models, the temperature $T$ is also given as a power law. \citep{idalin2004a} for example assumed an optically thin disk, and a main sequence-scaling of the stellar luminosity as $L\propto M_{\star}^{4}$ with stellar mass, so that
\beq\label{eqTr}
T(r)=280  {\rm K} \left(\frac{r}{\rm 1 AU}\right)^{-1/2} \left(\frac{M_{\star}}{\msun}\right).
\eeq
This however neglects  (a) that disks are optically thick (both radially and vertically) with opacity transitions at condensation fronts which can act as migration traps, (b) the effect of viscous heating, and (c) it does not include any temporal evolution, including the fact that stars are not yet on the main sequence during the presence of the gas disk. A certain improvement over such simple MMSN-like models are analytical disk models that take these effects into account, {as} for example the \citet{chambers2009} disk model that distinguishes between an inner viscously heated part and an outer irradiation-dominated part. 

A more complex, but still 1D approach is to solve the classical viscous evolution equation \citep{luest1952,lynden-bellpringle1974} for the surface density of the gas as a function of time $t$ and distance from the star $r$
\beq\label{eq:lyndenbellsigma}
\frac{\partial \Sigma}{\partial t}=\frac{1}{r}\frac{\partial}{\partial r}\left[3 r^{1/2} \frac{\partial}{\partial r}\left(r^{1/2}\nu \Sigma\right)\right]-\dot{\Sigma}_{\rm phot}(r)-\dot{\Sigma}_{\rm planet}(r).
\eeq
with a viscosity $\nu$ that is written in the $\alpha$-parameterization as $\nu=\alpha c_{\rm s} H$ with $c_{\rm s}$ the sound speed and $H$ the vertical scale height \citep{shakurasunyaev1973}. Besides the viscous evolution term, the effects of mass loss by photoevaporation \citep[e.g.,][]{alexanderpascucci2014} represented by $\dot{\Sigma}_{\rm phot}(r)$ and of gas accretion by the planets giving raise to the $\dot{\Sigma}_{\rm planet}(r)$ term are also to be included.

As an initial condition for this equation, the gas surface density is assumed to consist of a decrease close to the star due to the stellar magnetospheric cavity, a power-law in the main part, and an exponential decrease outside of a characteristic radius, as found in the analytical solution to the viscous accretion disk problem of \citet{lynden-bellpringle1974}. The initial gas surface density  is then
\beq\label{eq:initialsigma}
\Sigma_{\rm g}(t=0,r)=\Sigma_{\rm 0} \left(\frac{r}{\rm 1 AU}\right)^{p_{\rm g}} \exp\left[{-\left(\frac{r}{R_{\rm out}}\right)^{2+p_{\rm g}}}\right]\left(1-\sqrt{\frac{r}{R_{\rm in}}}\right).
\eeq 
In this equation, $R_{\rm out}$ is the ``characteristic'' (outer) disk radius, $R_{\rm in}$ the inner radius, and $p_{\rm g}$ the power law exponent. Observations indicate  $p_{\rm g}\approx -1$ \citep{andrewswilner2010}. The four parameters in this equation may be treated as  Monte Carlo random variables in a population synthesis.

Under the assumption that dust is converted early in the disk's evolution everywhere with full efficiency into planetesimals, the initial surface density of planetesimals $\Sigma_{\rm p}$ would be given as \citep{mordasinialibert2009a}
\beq\label{eqSigmad}
\Sigma_{\rm p}(t=0,r)=f_{\rm dg}\eta_{\rm ice} \Sigma_{\rm g}(t=0,r)
\eeq
where $f_{\rm dg}$ is the dust-to-gas ratio ($\approx$ the heavy element mass fraction $Z$), which is about 0.0149 in the Sun \citep{lodders2003}. It is another Monte Carlo variable, {representing the different metallicities of stars (see Fig. \ref{CMinitdist})}. Finally, $\eta_{\rm ice}$ reflects the reduction of the solid surface density at iceline(s). 

However, observations \citep[e.g.,][]{panichogerheijde2009} and theoretical results \citep[e.g.,][]{birnstielklahr2012} indicate that a significant radial redistribution of solids in the form of pebbles occurs. This may lead to more concentrated and steeper distributions of the solids \citep{kornetstepinski2001,birnstielandrews2014} than predicted by Eq \ref{eqSigmad}. In pebble-based models, the pebble surface density is calculated from the  radial flux of pebbles, which is in turn controlled by the production rate of pebbles from dust at the pebble production line \citep{bitschLambrechts2015}.

The temporal evolution of the gas disk is found by solving the aforementioned equation describing a viscous accretion disk including photoevaporation. In parameterized models like in \citet{idalin2004a}, one {uses} instead an equation of the form
\beq\label{eq:sigmataudecrease}
\dot{\Sigma}_{\rm g}(r)=-\frac{\Sigma_{\rm g}(r)}{\tau_{\rm disk}}+\dot{\Sigma}_{\rm phot}.
\eeq 
The first term on the right hand side leads to an exponential self-similar decay, while the second mimics the effects of photoevaporation. The characteristic disk timescale $\tau_{\rm disk}$ can again be treated as a Monte Carlo variable ({Sect. \nameref{sect:probdists}}). 

The surface density of solids decreases within the planet's feeding zone according to the amount of mass that the planet accretes, assuming that the surface density is uniform within the feeding zone \citep{thommesduncan2003}, i.e.,
\beq
\dot{\Sigma}_{\rm p} = - \frac{(3 \mstar)^{1/3}}{6\pi a_{\rm p}^2 B_L M_{\rm p}^{1/3}}\dot{M}_{\rm c}
\eeq
where $B_L$ is the width of the feeding zone in Hill spheres, $M_{\rm p}$ the planet's mass, $a_{\rm p}$ its semimajor axis, and $\dot{M}_{\rm c}$ the planet's planetesimals accretion rate. 

\subsubsection{Accretion of solids}\label{sect:accretionofsolids}
In planetesimal based models, the growth of the solid core with mass $M_{\rm c}$ is assumed to occur in the classical picture via the accretion of small background planetesimals. For this, a version of the  Safronov \citep{safronov1969} equation which gives the core accretion rate $\dot{M}_{c}$ is used: 
\beq\label{eq:safronov}
\dot{M}_{\rm c}=\Omega\Sigma_{\rm p} R_{\rm capture}^2 F_{\rm G}
\eeq
where $\Omega$ is the Keplerian frequency, $\Sigma_{\rm p}$ the mean surface density of planetesimals in the planet's feeding zone,
$R_{\rm capture}$ the capture radius which is in general larger than the core radius because of gas drag \citep{podolakpollack1988,mordasinialibert2006b}, and $F_{\rm G}$ is the gravitational focussing factor \citep{nakazawaida1989a,greenzweiglissauer1992}. It  depends {among other quantities} on the random velocities of the planetesimals {$v_{\rm pls}$ and would be given in the (idealized) two-body case as $1+(v_{\rm esc}/v_{\rm pls})^{2}$, where $v_{\rm esc}$ is the escape velocity from the protoplanet \citep{safronov1969}.} {The random velocities of smaller planetesimals are more strongly damped by nebular gas drag leading to a higher focussing factor in Eq. \ref{eq:safronov}. Furthermore, the drag-enhanced capture radii of the protoplanets in Eq. \ref{eq:safronov} is increased as well for smaller bodies, approaching very large radii for small particles as exemplified by pebble accretion.}

An insight into the dependencies of the core accretion rate on parameters can be obtained by considering the core accretion timescale $\tau_{\rm c}$ in 
\beq\label{dotMc}
\dot{M}_{\rm c}=\frac{M_{\rm c}}{\tau_{\rm c}}.
\eeq
Based on the work of \citet{kokuboida2002}, \citet{idalin2004a} derive an approximate expression for the accretion timescale  in the oligarchic growth regime. In the oligarchic regime, the random velocities of  the planetesimals is raised by viscous stirring by the protoplanet, while it is damped by gas drag during the presence of the gas disk {\citep{idamakino1993}. This regimes occurs after an initial runaway planetesimal accretion phase as soon as the protoplanets have grown to a size of 100-1000 km depending on orbital distance \citep{ormeldullemond2010}.} 

The accretion timescale in this regime is
\begin{multline}\label{eq:taucoreaccr}
\tau_{\rm c}=1.2\times10^{5} {\ \rm yr}\left(\frac{\Sigma_{\rm p}}{10\ {\rm  g \  cm}^{-2}}\right)^{-1}\left(\frac{a_{\rm p}}{1\ {\rm  AU}}\right)^{1/2}\left(\frac{M_{c}}{\mearth}\right)^{1/3}\left(\frac{M_{\star}}{M_{\odot}}\right)^{-1/6}  \times \\ \left[  \left(\frac{\Sigma_{\rm g}}{2400 {\rm   \ g \  cm}^{-2}}\right)^{-1/5} \left(\frac{a_{\rm p}}{1\ {\rm  AU}}\right)^{1/20} \left(\frac{m}{10^{18}\ {\rm  g}}\right)^{1/15}\right]^{2}.
\end{multline}
In this equation $\Sigma_{\rm g}$ is the gas surface density at the planet's position at $a_{\rm p}$, and $m$ is the mass of a planetesimal. 

We see from Eq. \ref{eq:taucoreaccr} that the growth is faster at smaller distances as the collisional growth scales with the orbital frequency leading to growth wave propagating outward. The timescale also increases as $M_{\rm c}^{1/3}$ as typical for the oligarchic regime, and decreases inversely proportional to $\Sigma_{\rm p}$. This faster growth, and the fact that cores can also become more massive at higher $\Sigma_{\rm p}$ \citep[e.g.,][]{kokuboida2012} explains why the core accretion theory predicts a higher number of giant planets at higher [Fe/H] (see Sect. \nameref{corrdiskprops}).

{In pebble based models  \citep[see, e.g., ][]{ormel2017}, the accretion of pebble sets in once the protoplanets have reached a size where the encounter with the pebbles transitions from the  ballistic to the settling regime. In the former, gas-drag effects are not relevant, whereas in the settling regime the encounter time is sufficiently long to allow the incoming particles to couple aerodynamically to the gas, and to sediment towards the protoplanet during the encounter. This leads to an efficient, gas-drag-aided accretion. This transition occurs at one AU when the protoplanets have reached a size of several hundred km, increasing with orbital distance \citep{visserormel2016}. }

\subsubsection{Accretion of gas}\label{sect:accretiongas}
Two approaches are used  in global models to calculate a protoplanet's gas accretion rate. The first more complex approach taken for example in the Bern model is to calculate the interior structure of the (proto)planets \citep{alibertmordasini2005,mordasinialibert2012b}. The planets' interior is modeled by integrating numerically the 1D spherically symmetric structure equations  which are the mass conservation, hydrostatic, energy conservation, and energy transport equations \citep{bodenheimerpollack1986}:
\begin{alignat}{2}\label{eq:internalstruct}
\frac{\partial m}{\partial r}&=4 \pi r^{2} \rho    &\quad  \quad \frac{\partial P}{\partial r}&=-\frac{G m}{r^{2}}\rho    \\
\frac{\partial l}{\partial r}&=4 \pi r^{2} \rho\left(\varepsilon -P \frac{\partial V}{\partial t} -\frac{\partial u}{\partial t}\right)             & \frac{ \partial T}{\partial r}&=\frac{T}{P}\frac{\partial P}{\partial r}\nabla(T,P)          
\end{alignat}
where $r$ is the radius measured from the planet's center, $m$ the enclosed mass, $P$ the pressure, $\rho$ the density,  and $G$  the gravitational constant. The gradient $\nabla$ depends on the process by which the energy is transported (radiation or convection). The energy equation is the only equation that is time $t$ dependent and controls the temporal evolution. In this equation, $V=1/\rho$ is the specific volume, $u$ the specific internal energy, $\varepsilon$ an energy source like impact or radiogenic heating, and $l$ is the intrinsic luminosity.

These structure equations are solved with different outer boundary conditions depending on the phase a protoplanet is in \citep{bodenheimerhubickyj2000,mordasinialibert2012b}. In the first so-called attached (or nebular) phase, the envelope transitions smoothly into the background nebula, such that the outer pressure and temperature are approximately equal to the local disk pressure and temperature. In this phase, the outer radius is given as the minimum of a fraction of the Hills sphere and the Bondi radius, and can thus be found if the planet's mass is known. Radiative cooling allows the gas in the protoplanetary envelope to contract. This (formally) results in an empty shell between the planet's outer edge of the envelope and the surrounding nebula. This is filled in by new nebular gas, allowing the envelope mass to increase. This means that during the attached phase, the gas accretion rate is regulated by the envelope's cooling (Kelvin-Helmholtz) timescale as found by solving the structure equations.

When the core reaches a mass of  about 10 $\mearth$, the contraction of the envelope becomes so rapid  that the protoplanetary disk can no {longer} supply gas at a rate sufficient to keep the envelope and disk in contact (runaway gas accretion leading to giant planet formation).  The planet's outer radius now detaches from the nebula and contracts rapidly, but still quasi-statically \citep{bodenheimerpollack1986} to a radius that is much smaller than the Hills sphere (about 1.5 - 5  $\rj$ depending on the entropy, \citealt{mordasinialibert2012b,mordasinimarleau2017}). In this second so-called detached (or transition) phase, the radius is free and found by solving the structure equations, while the gas accretion rate is given by processes in the protoplanetary disk and no {longer}  by the envelope's contraction. This disk-limited rate may be given by the Bondi accretion rate (\citealt{dangelolubow2008,mordasinialibert2012b})
\beq\label{eq:mdotbondi}
\dot{M}_{\rm e, Bondi}\approx\frac{\Sigma_{\rm g}}{H}\left(\frac{R_{\rm H}}{3}\right)^{3}\Omega
\eeq
where $\Sigma_{\rm g}$, $H$, $R_{\rm H}$, and $\Omega$ are the mean gas surface density in the planet's feeding zone, the disk's vertical scale height, the planet's Hill sphere radius, and the orbital frequency at the planet's position. It is also possible to calculate the planet's disk-limited gas accretion rate as a fraction  $f_{\rm lub}$ of the local viscous accretion rate in the disk, which is in equilibrium given as
\beq
\dot{M}_{\rm e, visc}=  f_{\rm lub} 3 \pi \nu \Sigma_{\rm g}.
\eeq
Hydrodynamical simulations \citep{lubowseibert1999} indicate that $f_{\rm lub}$ may be as high as 0.9  meaning that the planet accretes 90\% of the local gas flow through the disk. At higher masses, gap formation starts to reduce the gas accretion rate, leading to a reduction  which can be fitted as \citep{verasarmitage2004} 
\beq
f_{\rm va04}=1.668 \left(\frac{M_{\rm p}}{M_{\rm Jup}}\right)^{1/3}\exp\left(-\frac{M_{\rm p}}{1.5 M_{\rm Jup}}\right)+0.04.
\eeq

The second simpler approach used by other global models \citep[e.g.,][]{idalin2004a,hasegawapudritz2012,ndugubitsch2018} to calculate the gas accretion rate is based on fits for the KH-timescale. The gas accretion rate  due to the contraction of the envelope is approximated as 
\beq
\dot{M}_{\rm e, KH}=\frac{M_{\rm p}}{\tau_{\rm KH}}
\eeq
where the Kelvin-Helmholtz cooling timescale of the envelope is parameterized as  \citep{ikomanakazawa2000} 
\beq\label{tKH}
\tau_{\rm KH}=10^{p_{\rm KH}} {\ \rm yr} \left(\frac{M_{\rm p}}{\mearth}\right)^{q_{\rm KH}}\left(\frac{\kappa}{1 {\rm \ g \ cm}^{-2}}\right)
\eeq
where $p_{\rm KH}$ and $q_{\rm KH}$ are parameters that are obtained by fitting the accretion rate found with internal structure calculations like \citet{bodenheimerpollack1986,ikomanakazawa2000,mordasiniklahr2014}. For example, \citet{idalin2004a} used $p_{\rm KH}=9$ and $q_{\rm KH}=-3$ and neglected the influence of $\kappa$. \citet{mordasiniklahr2014} found $p_{\rm KH}=10.4$, $q_{\rm KH}=-1.5$, and $\kappa=10^{-2}$ g/cm$^{2}$. Once can see from Eq. \ref{tKH} that the accretion rate is a rapidly increasing function of mass, and that gas accretion becomes important once $\tau_{\rm KH}$ becomes comparable to, or shorter than, the disk lifetime.

\begin{figure}[h]
    \centering
    \includegraphics[width=\textwidth]{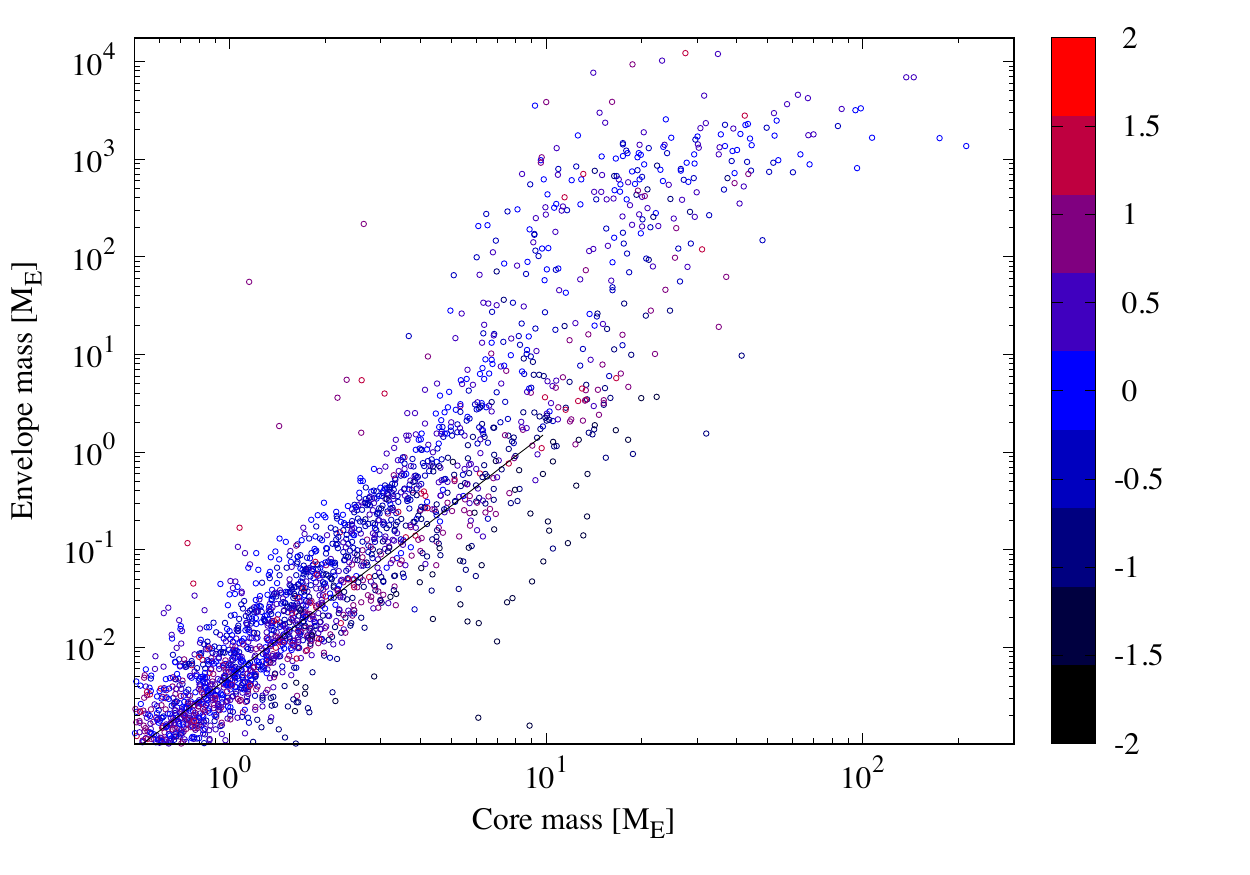}
   \caption{H/He envelope mass as a function of core mass found from solving the internal structure equations for the {synthetic} population discussed in Sect. \nameref{sect:results}. The relation is shown at the end of the formation phase when the gaseous disk{s} disperse. The colors show the planets' semimajor axis as $\log(a/\mathrm{AU})$. The black line scales as $\mcore^{2.5}$.}\label{McoreMenve}
\end{figure}

Compared to the method of solving directly the internal structure, the KH-method is computationally much simpler and more robust. But it cannot take into account how the gas accretion rate depends on (a) the (variable) luminosity of the core because of solid accretion, and (b)  the varying outer boundary conditions. {This means that it cannot easily recover the spread in associated envelope masses for a fixed core mass} visible in Fig. \ref{McoreMenve}. Furthermore it does not yield the planets' internal structure and thus radius and luminosity. But also the solution of the 1D hydrostatic structure equations is only an approximation as it neglects that protoplanetary envelopes are not closed strictly hydrostatic 1D systems, but that they can exchange gas with the surrounding disk in a hydrodynamic multi-dimensional manner \citep{ormelshi2015}. This can delay gas accretion through the advection of high entropy material \citep{Cimermankuiper2017}. {As it is typical for global end-to-end models to rely on low-dimensional approaches (1D or 1+1D) because of computational efficiency, these effects were not considered so far in population syntheses.}

Figure \ref{McoreMenve} shows the envelope mass as a function of core mass at the end of the formation phase in the population around 1 $\msun$ stars presented in Sect. \nameref{sect:results}. The{se} envelope masses were found by solving the {aforementioned 1D} internal structure equations {assuming a grain opacity in the protoplanetary atmospheres that is 0.003 times as large as the ISM grain opacities \citep{mordasiniklahr2014}, and by limiting the gas accretion rate in the disk-limited phase similarly to Eq. \ref{eq:mdotbondi} (see \citealt{mordasinialibert2012b} for details)}. At low masses, the envelope mass scales as $M_{\rm c}^{-q_{\rm KH}+1}$ (i.e.,  $M_{\rm c}^{2.5}$ in the simulation here, indicated by the black line, see also \citealt{mordasiniklahr2014}). Then, at a core mass of about 5-20 $\mearth$, gas accretion becomes rapid (runaway accretion), so planets move upwards nearly {vertically} to higher $M_{\rm e}$ and become giant planets. 

There are fewer planets in the intermediate mass range between about 10 to 100 $\mearth$. This is because the timescale to accrete this gas mass in runaway accretion is  shorter than the disk lifetime, so that it is unlikely that the disk disappears exactly at an intermediate moment/mass. This is the origin of the so called ``planetary desert'' \citep{idalin2004a}. It is weaker in the population here compared to the {original} \citet{idalin2004a} predictions due to the larger effective $q_{\rm KH}=-1.5$ instead of -3 as used by \citet{idalin2004a}, meaning that the gas accretion rate does not  increase as rapidly with mass, and due to the limits given by the Bondi  rate.  Regarding the termination of gas accretion, in this simulation, the disk-limited gas accretion decreases in time just because $\Sigma_{\rm g}$ in the feeding zone (Eq. \ref{eq:mdotbondi}) decreases because of disk evolution. Gas accretion is thus  terminated when the gas disk disappears.

\subsubsection{Orbital migration}\label{sect:orbitalmig}
The gravitational interaction of the gaseous disk and the embedded protoplanets results in the exchange of angular momentum \citep[for recent reviews see][]{kleynelson2012,baruteaubai2016}, which means that the planets change their semimajor axis, i.e., the undergo orbital migration \citep{goldreichtremaine1979,ward1986,linpapaloizou1986}. The angular  momentum transfer between disk gas and planets via torques leads in most cases to a loss of angular momentum for the planet which means inward migration. The angular momentum $J$ of a planet of mass $M_{\rm p}$ orbiting a star of mass $\mstar$ at a semimajor axis $a_{\rm p}$, and the migration rate $d a/dt$ given a total torque $\Gamma_{\rm tot}=d J/ d t$ are 
\begin{align}\label{eq:migrationtorque}
    J&=M _{\rm p}\sqrt{G \mstar a_{\rm p}} &  \frac{d a}{dt}&=2 a_{\rm p} \frac{\Gamma_{\rm tot}}{J}.
\end{align}
Other effects like planetesimal-driven migration \citep[e.g.,][]{levisonthommes2010,ormelida2012} or Kozai migration due to an external perturber \citep{Kozai1962,fabryckytremaine2007} can also modify the orbits, but were up to now not considered in population synthesis models. 

Disk-driven migration occurs in two types, Type I and Type II migration. Type I migration occurs if the planet's Hill sphere radius is smaller than the disk's vertical scale height and if the viscous torques are dominant compared to the gravity torques induced by the planet \citep{cridamorbidelli2006}, meaning that Type I migration applies to low-mass planets. Various descriptions of Type I migration have been derived in the literature. Early derivations \citep{tanakatakeuchi2002} assumed that the disk behaves {(locally)}  isothermal, and predicted rapid inward migration. Later, more realistic calculations \citep[e.g.,][]{baruteaumasset2008,casolimasset2009,paardekooperbaruteau2010,kleybitsch2009} directly modeled the cooling behavior of the disk gas. They showed that there are several sub-types of Type I migration (locally isothermal, adiabatic, (un)saturated corotation torque) that can be identified by considering an number of timescales \citep{dittkristmordasini2014}. An important finding is that {for non-isothermal Type I migration,} in some parts of the disk  outward migration can occur. {Therefore, }there are special locations {like condensation fronts} where the torque vanishes {because of the associated opacity transitions}. Such locations can act as traps for migrating planets \citep{lyrapaardekooper2010,hasegawapudritz2011a,sandorlyra2011b,kretkelin2012b}, and can serve as locations of efficient planetary growth \citep[e.g.,][]{hornlyra2012,hasegawapudritz2012}. 

To calculate the torque causing a planet's migration, one needs among other quantities like the gas surface density the local power law exponent of the disk temperature $p_{\rm T}$ and of the gas surface density $p_{\rm \Sigma}$, which are yielded by the disk model, underlining its importance. The migration timescale in the isothermal approximation used by \citet{idalin2008c} is given as
\beq\label{eq:tautypeI}
\tau_{\rm type I}=\frac{1}{2.728+1.082 p_{\rm \Sigma}}\left(\frac{c_{\rm s}}{a_{\rm p}\Omega}\right)^{2}\frac{\mstar}{M_{\rm p}}\frac{\mstar}{a_{\rm p}^{2} \Sigma_{\rm g}} \Omega^{-1}
\eeq
and the migration rate is then
\beq
\dot{a}_{\rm p}=-\frac{a_{\rm p}}{\tau_{\rm type I}}
\eeq
which shows that migration speeds up a  planet grow more massive. In a more recent analysis, \citet{paardekooperbaruteau2010} found that the total torque $\Gamma_{\rm tot}$ resulting from summing up the contributions from the inner and outer Lindblad torques plus the corotation torque can be expressed in an equation of the form 
\beq\label{eq:typeIgamma}
    \Gamma_{\rm tot}=\frac{1}{\gamma}(C_{0}+C_{1}p_{\rm \Sigma}+C_{2}p_{\rm T}) \Gamma_{0} 
    \eeq
 with
    \beq
    \Gamma_{0}  =\left(\frac{q}{h}\right)^{2}\Sigma_{\rm g} a_{\rm p}^{4}\Omega^{2 }
\eeq
where $\gamma$ is the ratio of the heat capacities, $h=H/a_{\rm p}$ the local disk aspect ratio, $q=M_{\rm p}/\mstar$, $\Sigma_{\rm g}$ the gas surface density at the planet's position, and $\Omega$ its Keplerian frequency. The constants $C_{i}$ depend on the Type I sub-regime. {Their numerical values are listed for example in \citet{dittkristmordasini2014}}. 

Type II migration occurs if the angular momentum injection rate of the planet into the disk is so large that it carves a gap into the gas disk around  its location. For global models, several different descriptions of Type II migration were considered in the literature: {\citet{idalin2004a} consider the angular momentum transfer rate in a viscous accretion disk without planets (the viscous torque or ``couple'' in the terminology of  \citealt{lynden-bellpringle1974}) and assume that planets in the type II migration regime act as relays that transmit angular momentum also at this rate across their gap via tidal torques. Inserting the viscous torque into Eq. \ref{eq:migrationtorque}, the} Type II migration rate is  given as
\beq
\dot{a}_{\rm p}=3 \ {\rm sign}(a_{\rm p}- R_{\rm m})  \alpha \frac{\Sigma_{\rm g,m} R_{\rm m}^{2}}{M_{\rm p}}\frac{\Omega_{\rm m}}{\Omega}\left(\frac{H_{\rm m}}{a_{\rm p}}\right)^{2}\Omega_{\rm m}
\eeq
where quantities with the subscript m are evaluated at the radius of maximum viscous couple (or velocity reversal, i.e., where the disk changes from accreting to decreting, see \citealt{lynden-bellpringle1974}). The position of $R_{\rm m }$ can either be estimated as in \citet{idalin2004a}
\beq
R_{\rm m }= 10 \ {\rm AU}\exp\left(\frac{2 t}{5 \tau_{\rm disk}}\right).
\eeq
or results automatically from  solving the evolutionary equation for the gas surface density (Eq. \ref{eq:lyndenbellsigma}).

The Type II migration description of \citet{alibertmordasini2005} assumes that a planet follows the motion of the gas except for the case that the planet is massive compared to the local disk mass, when the planet is assumed to slow down because of its inertia \citep{alexanderarmitage2009}. The migration rate is thus given as
\beq\label{eq:dotaown}
\dot{a}_{\rm p}= \ u_{\rm r} \ {\rm min}\left(1, \frac{ 2 \Sigma_{\rm g} a_{\rm p}^{2}}{M_{\rm p}}\right)
\eeq
where $u_{\rm r}$ is the local radial velocity of the accreting gas which is in equilibrium given as $ 3 \nu/ (2 a_{\rm p})$.  A more realistic  yet computationally still feasible approach for a population synthesis (i.e., a 1D approach) is to employ the impuls approximation \citep{linpapaloizou1986a} to estimate the Type II migration, as for example done in \citet{colemannelson2014}. 

Given recent results \citep[e.g.,][]{duffellhaiman2014,durmannkley2015} questioning the classical conception that planets in Type II migration simply follow the viscous evolution of the disk{, but that their migration rate is entirely given by the torques,} make it likely that migration models will undergo significant modifications in the future. The same is true for Type I migration, where new effects like an additional ``heating'' torque resulting from the protoplanet's accretional luminosity counteracts inward migration \citep{benitez-llambaymasset2015}, or a ``dynamic'' corotation torque \citep{paardekooper2014,pierens2015} that results from the fact that the relative motion of gas and a migrating planet can lead to a feedback (usually, torques are measured for planets at fixed positions). This can  lead  to outward migration as well.

\subsubsection{N-body interactions}
The concurrent formation of several protoplanets in a protoplanetary disk affects the growth history of the protoplanets in multiple ways:
the protoplanets compete for the accretion of gas and solids, increase the velocity dispersion of the planetesimals potentially reducing the solid accretion rate of neighboring protoplanets, alter the surface density of planetesimals and for giant planets of the gas, {and reduce the radial flux of pebbles}. The gravitational interaction between the protoplanetes leads {in the case of} insufficient damping by the gas disk to the excitation of the eccentricities, resulting in the alteration of the orbits, collisions, and ejections. Orbital {migration} is affected as well, since the planets can capture into mean motion resonances and migrate together as resonant convoys. For specific parameters, this can even invert the direction of migration \citep{massetsnellgrove2001} and lead to the outward migration of two giant planets, as invoked for the ``grand tack'' scenario in the solar system \citep{walshmorbidelli2011}.

As discussed {above}, all early population synthesis models used the one-embryo-per-disk approach, which was one of the most important limitation of the first generation of the models, in particular for low-mass planets as they  usually occur in multiple systems, often in compact configurations \citep[e.g.,][]{mayormarmier2011}. This means that they likely influenced each other during formation.
 
This limitation was  addressed in  \citet{idalin2010,idalin2013} and \citet{alibertcarron2013}. In the Bern model an explicit N-body integrator was added ``on top'' of the existing sub-models that calculates the N-body interactions and collisions of the concurrently forming protoplanets. In order to keep the computational time sufficiently low for population syntheses, about 20-50 low-mass embryos (0.01-0.1 $\mearth$) are put into each disk. They interact via the usual Newtonian N-body forces written in the heliocentric system,
\beq
\ddot{ \vec{r}}_i  =  - G \left(M_* + m_i \right) \frac{\vec{r_i}}{r_{i}^{3}} - G \sum_{j=1,j\neq i}^{n} {m_j \left\{ \frac{\vec{r_i}-\vec{r_j}}{\left| \vec{r_i}-\vec{r_j} \right|^3} + \frac{\vec{r_j}}{r_{j}^{3}} \right\} }
\eeq
with $i = 1,2,3 \dots N$ the planet index, $\vec{r_i}$ and $m_i$  the heliocentric position and mass of planet $i$, and $M_*$ the mass of the central star. The consequences of the gravitational interaction with the gas disk (orbital migration and damping of eccentricities and inclinations) are entered as additional forces into the integrator \citep{cresswellnelson2008}.

A different approach was taken by \citet{idalin2010} and \citet{idalin2013}, who developed a new semi-analytical approach to describe the gravitational interactions of several protoplanets in a statistical way based on orbit crossing timescales{, including the effect of resonant capture for migrating planets}. The advantage of this approach is a computational cost that is orders of magnitude lower than the direct N-body integration{, while still yielding distributions of the eccentricities and semimajor axes of interacting planets that agree well with the direct N-body simulations.}

\subsection{Probability distribution of disk initial conditions}\label{sect:probdists}
The second central ingredient for a population synthesis calculation are sets of initial conditions (see Fig. \ref{CMpopsynthworkflow}). These sets of initial conditions are drawn in a Monte Carlo way from probability distributions. These probability distributions represent the varying properties of protoplanetary disks and are derived as closely as possible from results of disk observations, or, if the quantities are not observable, from theoretical arguments. Typically, there are at least four Monte Carlo variables employed \citep{idalin2004a,mordasinialibert2009a}:

\begin{figure}[h]
    \centering
        \begin{minipage}{0.49\textwidth}
        \includegraphics[width=\textwidth]{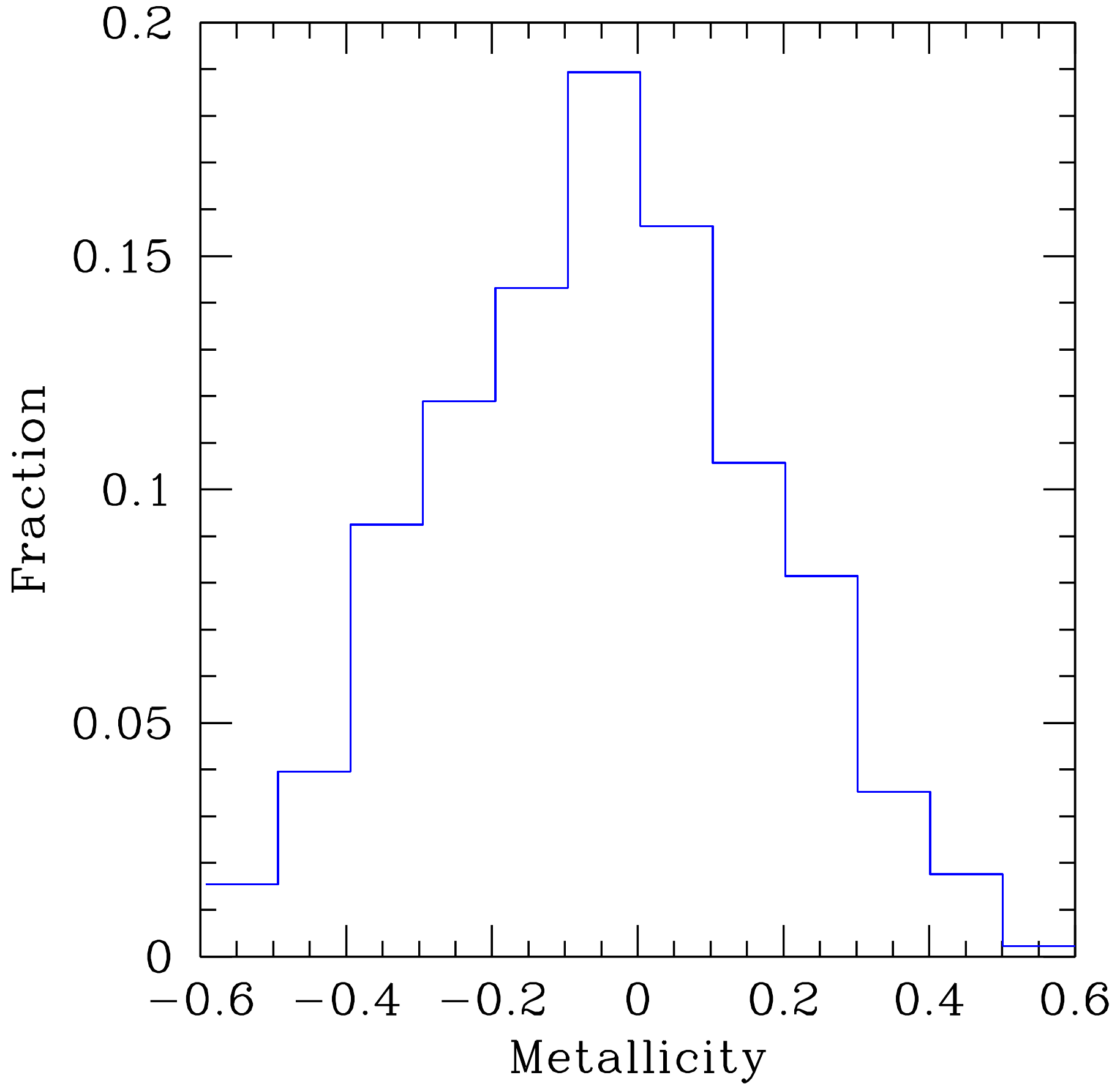}
	  \end{minipage}\hfill
        \begin{minipage}{0.47\textwidth}
        \includegraphics[width=\textwidth]{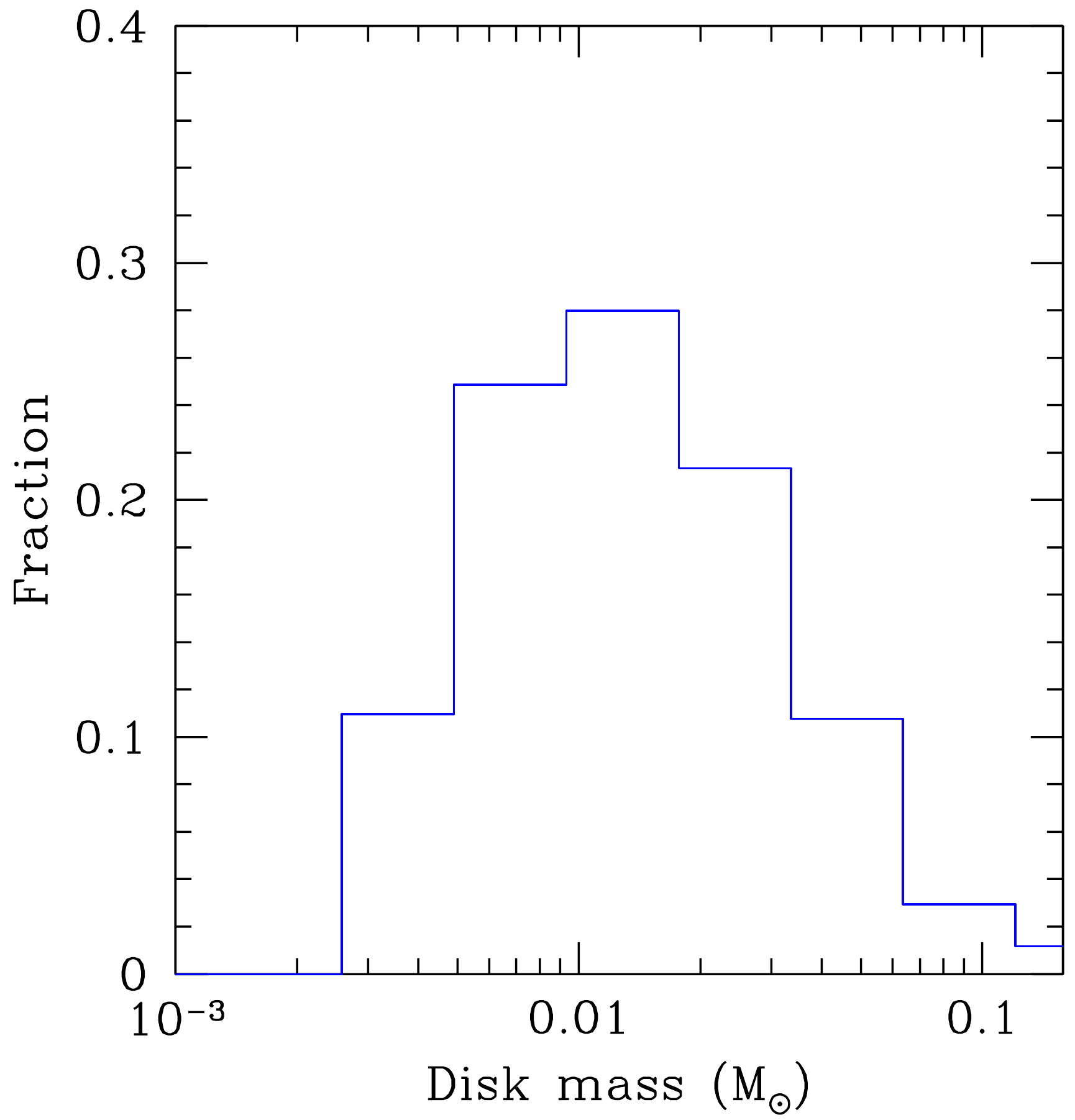}
         \end{minipage}\hfill
        \begin{minipage}{0.49\textwidth}
       \includegraphics[width=\textwidth]{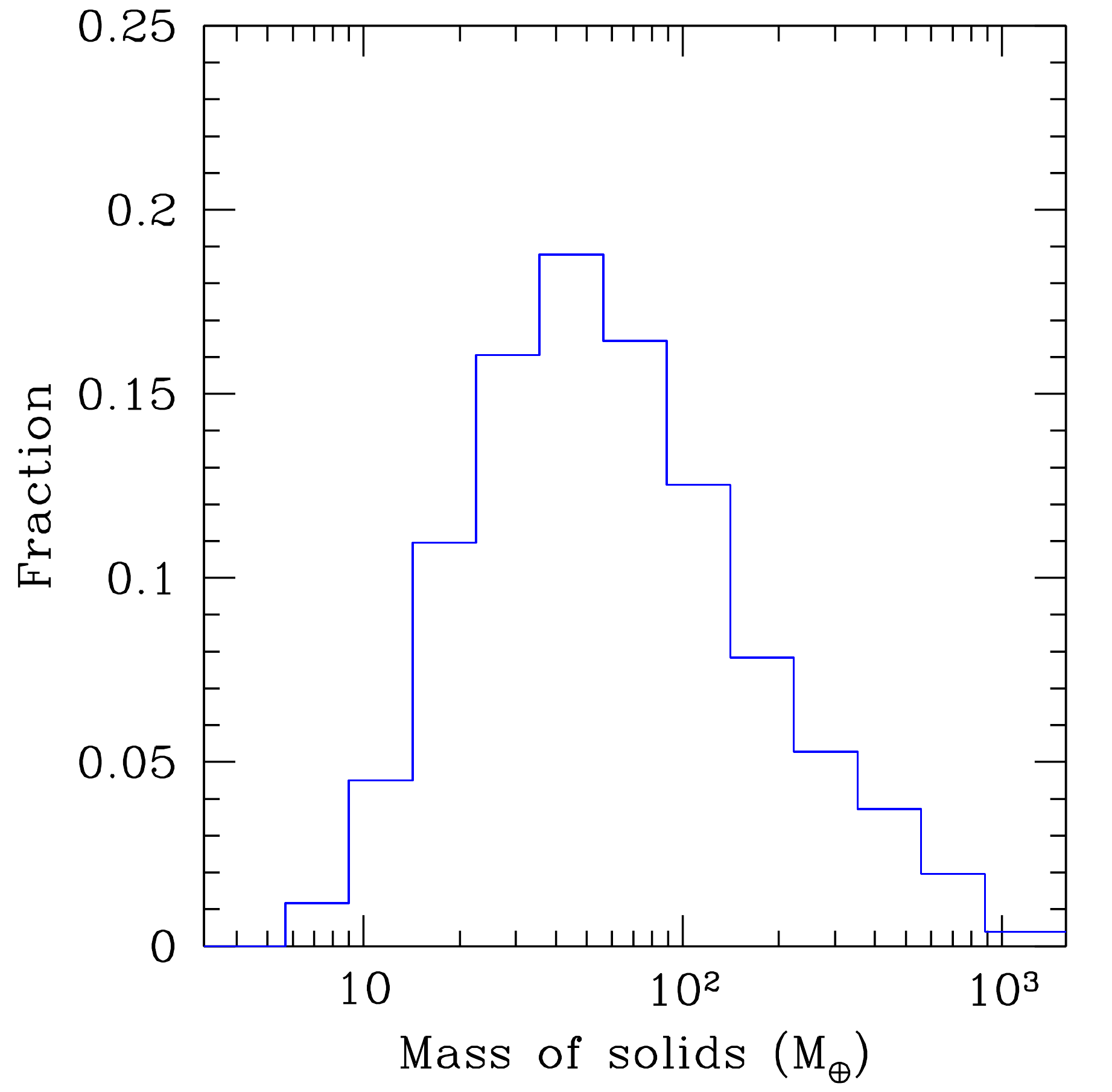}
         \end{minipage}\hfill
       \begin{minipage}{0.49\textwidth}
    \includegraphics[width=\textwidth]{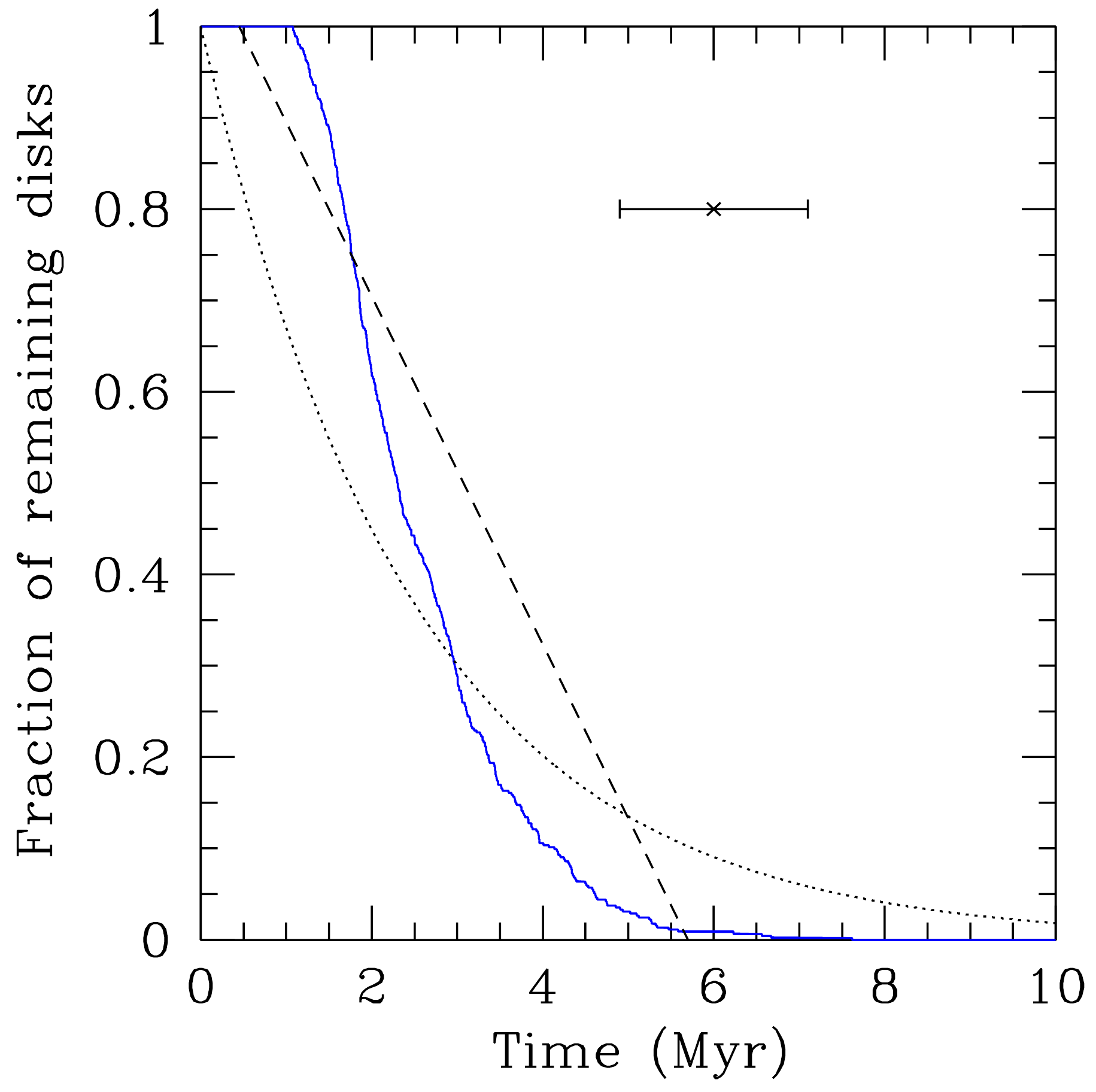}
         \end{minipage}\hfill
   \caption{Distributions of initial conditions for disk around 1 $\msun$ stars. Top left: Metallicity. Top right: initial disk gas mass. Bottom left: initial content of planetesimals. Bottom right: lifetime of synthetic disk (blue). The black solid and dotted lines show observationally determined lifetimes by \citet{haischlada2001} and \citet{mamajek2009}, respectively. The horizontal bars shows the typical observational age uncertainty.}\label{CMinitdist}
\end{figure}

\begin{enumerate}
\item \textit{The metallicity and dust-to-gas ratio} It is  usually assumed that the bulk metallicity is identical in the star and its protoplanetary disk. Then, the disk metallicity [M/H] can be modeled as a normal distribution as observed spectroscopically in the photosphere of solar-like stars in the solar neighborhood, with $\mu=-0.02$ and $\sigma$=0.22 \citep{santosisraelian2005}.  The [M/H] is converted into a disk dust-to-gas ratio {(Eq. \ref{eqSigmad})} via $f_{\rm dg}=f_{\rm dg,\odot} 10^{\mathrm{[M/H]}}$, with a solar $f_{\rm dg,\odot}$ of about 0.01 to 0.02 \citep{lodders2003}. Together with the initial disk gas mass and the locations of icelines, $f_{\rm dg}$ sets the amount of solids (dust, pebbles, planetesimals) available in the disk for planet formation.
\item \textit{The initial disk gas mass} The concept of an ``initial'' disk mass is of course questionable as it results from the dynamical collapse of a molecular cloud core \citep{shu1977,huesoguillot2005}, but it could be associated with the disk's mass at the moment when the main infall phase has ended, and no self-gravitational instabilities occur any more. Stability arguments \citep{shutremaine1990}, the inferred mass of the MMSN \citep{weidenschilling1977,hayashi1981}, and  observations of protoplanetary disk \citep{andrewswilner2010,manararosotti2016} point towards disk masses of about 0.1 to 10\% of the star's mass. The disk masses seem to be distributed roughly log-normally with a mean around 1\% of the star's mass \citep{mordasinialibert2009a}, but one should note that this distribution is poorly known.

\item \textit{The disk lifetime} The observations of IR and UV excesses of young stars indicate that the fraction of stars with protoplanetary disks decreases on a timescale of 1-10 Myr, with a mean lifetime of about 3 Myr \citep{haischlada2001,mamajek2009}. In a global model, this timescale can either be set directly in Eq. \ref{eq:sigmataudecrease}, or is used to find a distribution of  photoevaporation rates (Eq. \ref{eq:lyndenbellsigma}) that lead together with viscous accretion to a distribution of  lifetimes of the synthetic disks that agrees with observations. 
\item \textit{The initial starting positions of the embryos} Based on the finding of N-body simulations that oligarchs emerge with relative spacings of a few Hill spheres \citep{kokuboida2000}, a distribution of the starting embryos that is uniform in the $\log$ of the semimajor axis is usually used. It is also possible to arrange the embryos such that they ``fill'' the disk  taking into account the asymptotic planetesimal isolation mass \citep{idalin2010}. {In the trapped evolution models of \citet{hasegawapudritz2011a,cridlandpudritz2016} embryos rapidly move into traps, so that it is the locus and movement of the traps that effectively gives the formation locations.}
\end{enumerate}

Other quantities that may also be treated as Monte Carlo variables are for example the quantities describing the initial radial distribution of the gas and solids in Eq. \ref{eq:initialsigma}.  Other important parameters of the global models like the stellar mass, the planetesimals size, or -for viscous accretion disks- the $\alpha$ viscosity parameter \citep{shakurasunyaev1973} are usually kept constant for one synthetic population, but are varied across different populations to understand their statistical impact in parameter studies \citep[e.g.,][]{mordasinialibert2009b}. 

Figure \ref{CMinitdist} shows the distributions of  the disk (and stellar) metallicities, initial gas disk masses, the mass of planetesimals initially contained in the disks obtained with $f_{\rm dg,\odot}=0.02$, and the disk lifetimes. These are the initial conditions for the population synthesis described below in Sect. \nameref{sect:results}, containing 504 stars with 1 $\msun$. Note that in this model, the disk lifetime is not directly set, but results from the combined action of viscous accretion and an appropriately chosen distribution of photoevaporation rates.

\section{Results}\label{sect:results}
To illustrate what can be obtained from modern population synthesis calculations, we present in the following sections results from the latest generation of the Bern model. {The underlying global formation and evolution model used here is very similar to the model published in \citet{alibertcarron2013} regarding the N-body interactions of the protoplanets, and to the model published in \citet{mordasinialibert2012c} regarding the internal structure and long-term thermodynamic evolution (cooling, contraction, envelope evaporation) of the planets. In the new simulations presented here, these two aspects are now combined. This makes it possible to predict not only the orbital elements and masses, but also the radii and luminosities of planets in multi-planet synthetic systems. As  differences to the two previously published models, the} evolution of the star is now also considered {via} the Pisa stellar evolution tracks   \citep{dellomodarmeealle2012}{, and the Mercury N-body integrator \citep{chambers1999} is now employed}.  {As in previous models \citep[e.g.,][]{mordasinivanboekel2016}, the location of the  icelines is calculated with the initial disk structure and remains static in time under the assumption that efficient planetesimal formation happens early, and that for the 300-m planetesimals, drift is not very important, which should be the case at least in the outer nebula \citep[e.g.,][]{pisooeberg2015}. Clearly, this is a strong assumption. } 

\subsection{Initial conditions and parameters}\label{sect:initialconditionssynthesis}
Each system initially contains  20 {planetary} embryos with a starting mass of 0.1 $\mearth$. {These planetary seeds are} distributed randomly according to a log-uniform distribution between 0.05 and 40 AU. {Because of the influence of the initial condition, results concerning synthetic planets that are not clearly more massive than 0.1  $\mearth$ should be  regarded with caution.} The stellar mass is in all cases 1 $\msun$, and 504 star-disk systems are simulated. The formation phase of the systems was simulated during 10 Myr, during which the disks of gas and solid evolve, while the planets accrete mass, migrate, and interact and collide via the N-body integrator. Afterwards, the thermodynamic long-term evolution was calculated for 10 Gyr. During this later phase, the planets' mass is constant except for atmospheric escape and no dynamical interactions occur.

The initial gas surface density follows the profile in Eq. \ref{eq:initialsigma}, while the initial planetesimal follows a steeper profile $\propto r^{-1.5}$ as in the MMSN and a outer exponential radius that is half as large as the one for the gas (in Eq. \ref{eq:initialsigma}) to account for the inward drift of dust \citep{kornetstepinski2001,birnstielandrews2014} and the more concentrated distributions resulting from planetesimal formation \citep{drazkowskaalibert2016}. The planetesimal size is 300 meters, and the $\alpha$ viscosity parameter is 0.002. The grain opacity in the protoplanetary atmospheres during formation is reduced to 0.003 the ISM grain opacity \citep{mordasiniklahr2014}. During evolution, atmospheric opacities of a condensate-free solar-composition gas are assumed \citep{freedmanlustig2014}.

\subsection{Formation tracks}
Before discussing the statistical results, we present simulations obtained with the global model for one specific system, as the phenomena found in one system often help to understand the statistical results.

\begin{figure}[h]
    \centering
    \includegraphics[width=\textwidth]{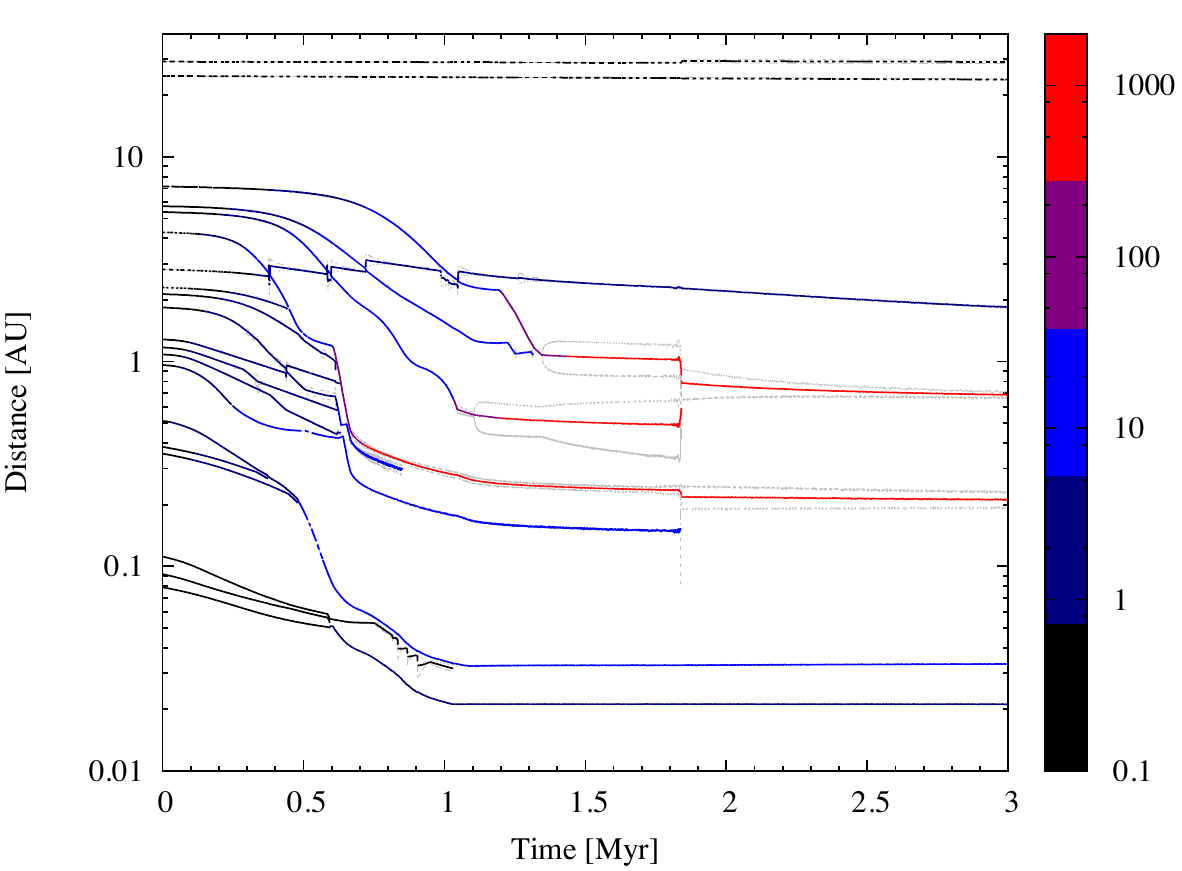}
   \caption{Inward migration, growth, and dynamical interaction in a synthetic system containing initially 20 planetary embryos of 0.1 $\mearth$ {The system is taken from the population synthesis described at the beginning of this section. The tracks of the planets} in the time versus orbital distance plane {are shown}. The black-blue-red lines show the planets' semimajor axes, with the color code representing the planets' mass in Earth masses. The grey lines show the apocenter and pericenter. Lines end when the corresponding protoplanet was either accreted by another  body{, or ejected because of dynamical interactions.} }\label{astartafinal}
\end{figure}

Figure \ref{astartafinal} illustrates the effect of planetary growth, N-body interaction, and orbital migration (Type I and II) for a system {taken from} the population described above. The effects of general inward migration, resonant capture, collisions, eccentricity excitation by planet-planet interaction, as well as eccentricity damping because of the gas disk can be seen. Starting from 20 embryos that are interacting via the N-body integrator, the system in the end contains 2 giant planets, a hot Neptunian planet, and 1 inner and 2 outer low-mass planets. 

\begin{figure}[h]
    \centering
    \includegraphics[width=\textwidth]{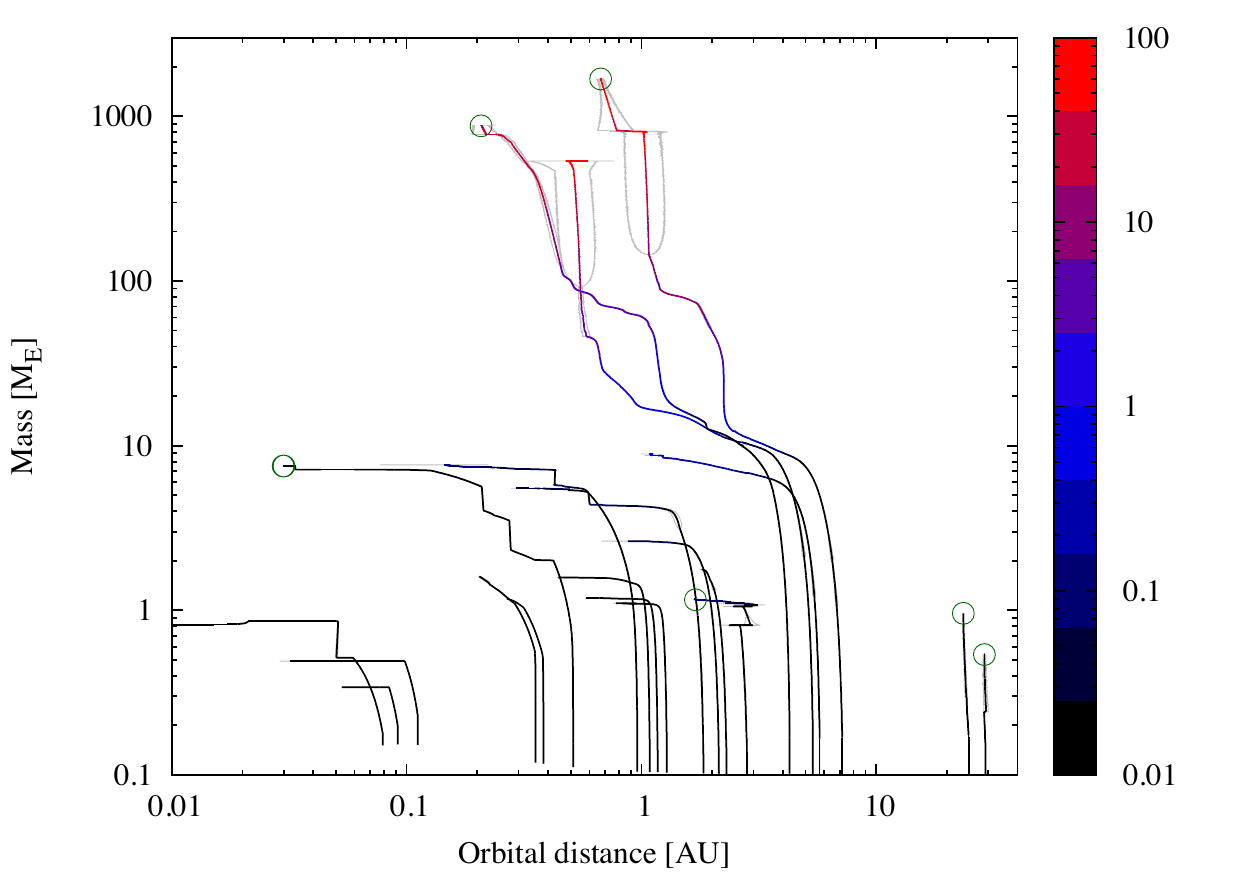}
   \caption{Growth tracks in the distance-mass plane in the same synthetic system as in Fig. \ref{astartafinal}.  The colored lines  show the semimajor axis, color coding the ratio of the H/He envelope mass relative to the to core mass $M_{\rm env}/M_{\rm core}$. Gray lines show the apocenter and pericenter distances. Open circles show the final position of the remaining planets. }\label{aMt111}
\end{figure}

{Orbital migration reduces the orbital distance for several planets by up to a factor 10. The existence of Type I migration traps at opacity transitions (Sect. \nameref{sect:orbitalmig}), the  slowing down of  Type II migration because of the giant planets' inertia (Eq. \ref{eq:dotaown}), and the finite disk lifetime still prevent the planets from falling into the star.}

Figure \ref{aMt111} shows growth tracks in the distance-mass plane in the same synthetic system as in Fig. \ref{astartafinal}. Several effects can be seen: at the beginning, the accretion timescale of solids is much shorter than the migration timescale, leading to nearly vertically rising tracks. With increasing mass, the solid accretion timescale (Eq. \ref{eq:taucoreaccr}) becomes longer, while the migration timescale becomes shorter (Eq. \ref{eq:tautypeI}), so the the planets start to migrate inwards at nearly constant mass once they have grown to about 5-10 $\mearth$. Some very low-mass planets are also captured into MMRs and pushed inwards by more massive protoplanets.

Three protoplanets grow so massive that they trigger runaway gas accretion occurring when $\mcore\approx\menv\approx10 \mearth$ as visible from the color code. During gas runaway, the growth tracks are again nearly vertical. Finally, the N-body interaction between the three giant planets increases their eccentricities {until their orbits overlap, as visible from the gray lines in Fig. \ref{astartafinal} and \ref{aMt111}}. {At about 1.9 Myr, this} lead{s} to the ejection of one {giant} planet {that was located at about 0.5 AU between the two surviving ones.}

\subsection{Diversity of planetary system architectures}\label{sect:diversityofplanetarysystemarchitectures}

The specific outcome in the simulation shown in Fig. \ref{astartafinal} and \ref{aMt111} depends obviously on the initial conditions, and  only shows one possible realization. To illustrate the architecture of planetary systems resulting from the global model and the distributions of initial conditions described in Sect. \nameref{sect:probdists}, we show in Fig. \ref{mamany} the final mass-distance diagram of 23 selected synthetic planetary systems. The systems are ordered from bottom left to top right according to increasing metallicity [M/H]. Note that the systems were selected by hand to display the diversity of architectures, and give the incorrect impression that systems with giant planets are common. This is not the case: only about 18 \% of all systems have a giant planet (see Sect. \nameref{sect:freqsynth}). {To first order, t}he systems can be {split} in three classes:

 \begin{figure}[h]
    \centering
    \includegraphics[width=0.99\textwidth]{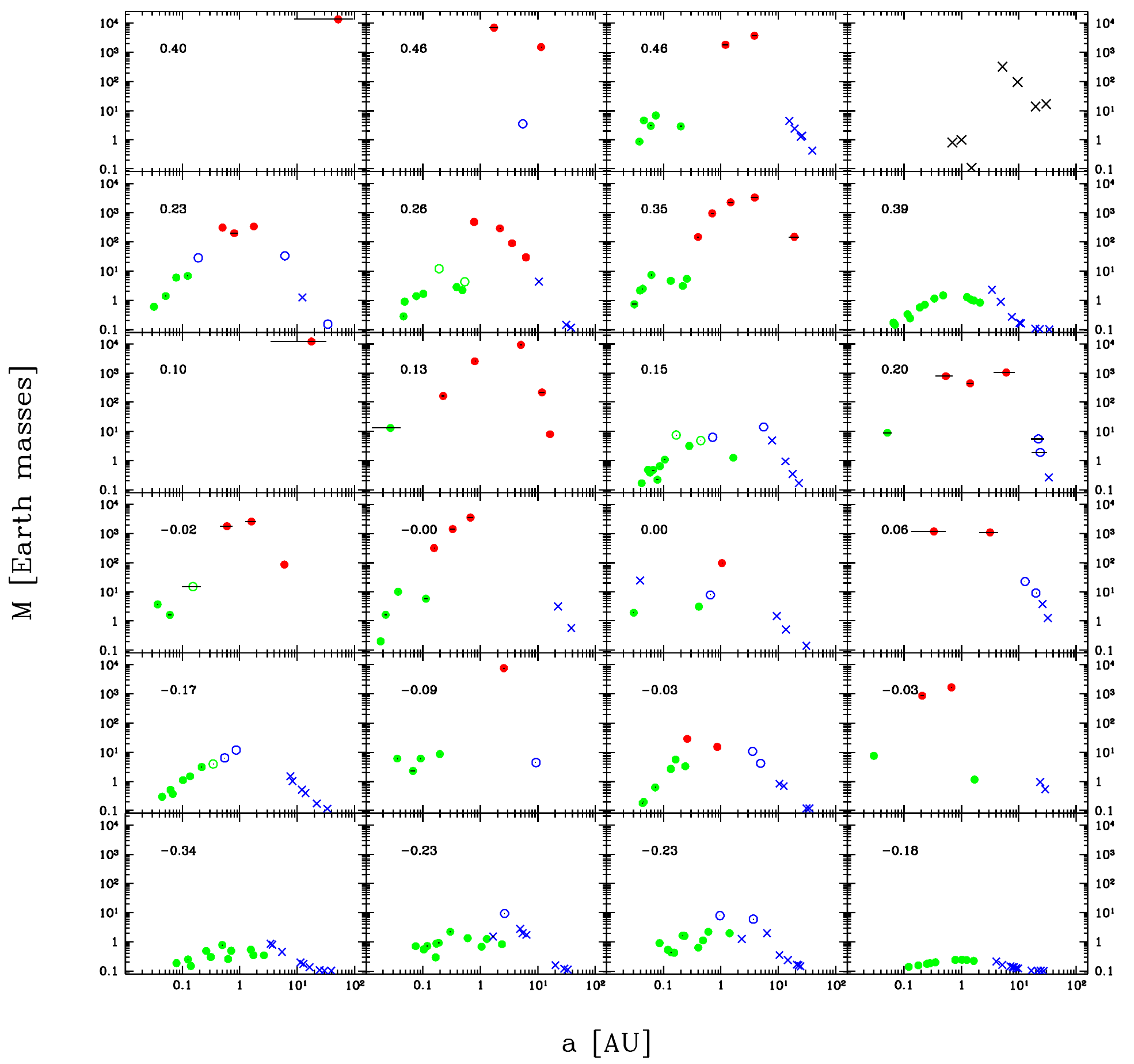}
   \caption{23 selected synthetic planetary systems in the mass-semimajor axis plane for 1 $\msun$ stars. {The system is taken from the population synthesis described at the beginning of this section.} Systems are ordered according to increasing [M/H], the number in each panel.   Red points are giant planets with $\menv/\mcore>1$. Blue symbols are planets that have (partially) accreted volatile material (ices) outside of the iceline(s), while green symbols have only accreted refractory solids. Open green and blue circles have 0.1$\leq\menv/\mcore\leq1$. Filled green points and blue crosses have $\menv/\mcore\leq0.1$. The black horizontal bars go from $a-e$ to $a+e$.  The top right panel is the solar system {for comparison}.}\label{mamany}
\end{figure}

(1) The large majority of the systems are similar to those visible in the bottom left corner: they only contain low-mass planets, with masses of 0.1-10 $\mearth$. These are systems where the disk properties are such (low surface densities of gas and solids, short disk lifetime) that only  little growth occurs during the first 10 Myr. Not much orbital migration and dynamical interaction has occurred because of the low planetary masses. Note that further growth on long timescales after 10 Myr is neglected in the model. This could first lead to further accretion, and second it could reduce the number of planets by giant impacts (in these systems the number of final planets is close to the initial number of embryos). The systems have a simple compositional transition from rocky to icy with increasing distance. The more massive planets ($\sim5-10\mearth$) have accreted gas envelopes comparable to Uranus and Neptune, otherwise little gas accretion has occurred (and was partially lost after formation by atmospheric escape which is modeled as described in \citealt{jinmordasini2014}). This type of system is preferentially forming at subsolar [M/H], but note that they also exist at high [M/H] (for example the [M/H]=0.15 and 0.39 systems). This makes clear that all four Monte Carlo variables play an important role in determining the outcome of the formation process.

(2) Already much less common are systems with giant planets and low-mass planets. Some contain only rocky low-mass planets inside of the giant planets (the [M/H]=-0.02 system), some only icy planets outside of them (the [M/H]=0.06 system). Some are  also reminiscent of the solar system and contain both inner terrestrial planets and outer icy planets (see the [M/H]=-0.00 system). Such a small-large-small arrangement is the classical outcome for collisional growth from planetesimals with little orbital migration (or redistribution of solids in general):  inside, the low availability of solids prevents much growth, while outside it is the long growth timescale that keeps planet masses low. The sweet spot for giant plant growth is a region outside of the water iceline. The architecture of the giant planets varies significantly:  the number {of giant planets in a system} varies from 1 to 5; in some systems the giants' mass increases with distance, in others it increases, and in some it is fairly constant; the eccentricities also range from near-zero value in many cases to higher values of about 0.2. 

However, despite the diversity, there is one common characteristic that distinguishes almost all synthetic systems from the solar system: the innermost giant planet is clearly closer-in than Jupiter, namely at about 1 AU or even less. This is an {intriguing} result. For the solar system, the ``grand tack'' model \citep{walshmorbidelli2011} suggests that Jupiter (and Saturn) also migrated to about 1.5 AU, to then migrate outward because of the Masset-Snellgroove effect occurring for resonantly coupled giants \citep{massetsnellgrove2001}. The way Type II migration is calculated in the model  here from the gas' radial velocity (Eq. \ref{eq:dotaown}) does not allow such outward migration. It will be interesting to see whether the inclusion of {more realistic migration models (see Sect. \nameref{sect:orbitalmig}) will} also lead to more distant {synthetic} giant planets.

(3) In a small fraction of systems, only one massive giant planet remains at the end. Such systems form in metal-rich and massive disk where several giant planets form in vicinity, leading to violent planet-planet scattering. In Plot \ref{mamany}, in the [M/H]=0.10 and 0.40 systems such giant planets with masses exceeding 30 $\mj$ can be seen with semimajor axes of 20-50 AU and high eccentricities. This gives them pericenter distances of less than 10 AU where the  scattering occurred and apocenter distances approaching in one case  of about 100 AU. They may be detectable by direct imaging. In such systems, all other planets were either accreted, sent into the star, or ejected. 

\subsection{The $a-M$ distribution}\label{sect:amdistribution}
\begin{figure}[h]
    \centering
    \includegraphics[width=\textwidth]{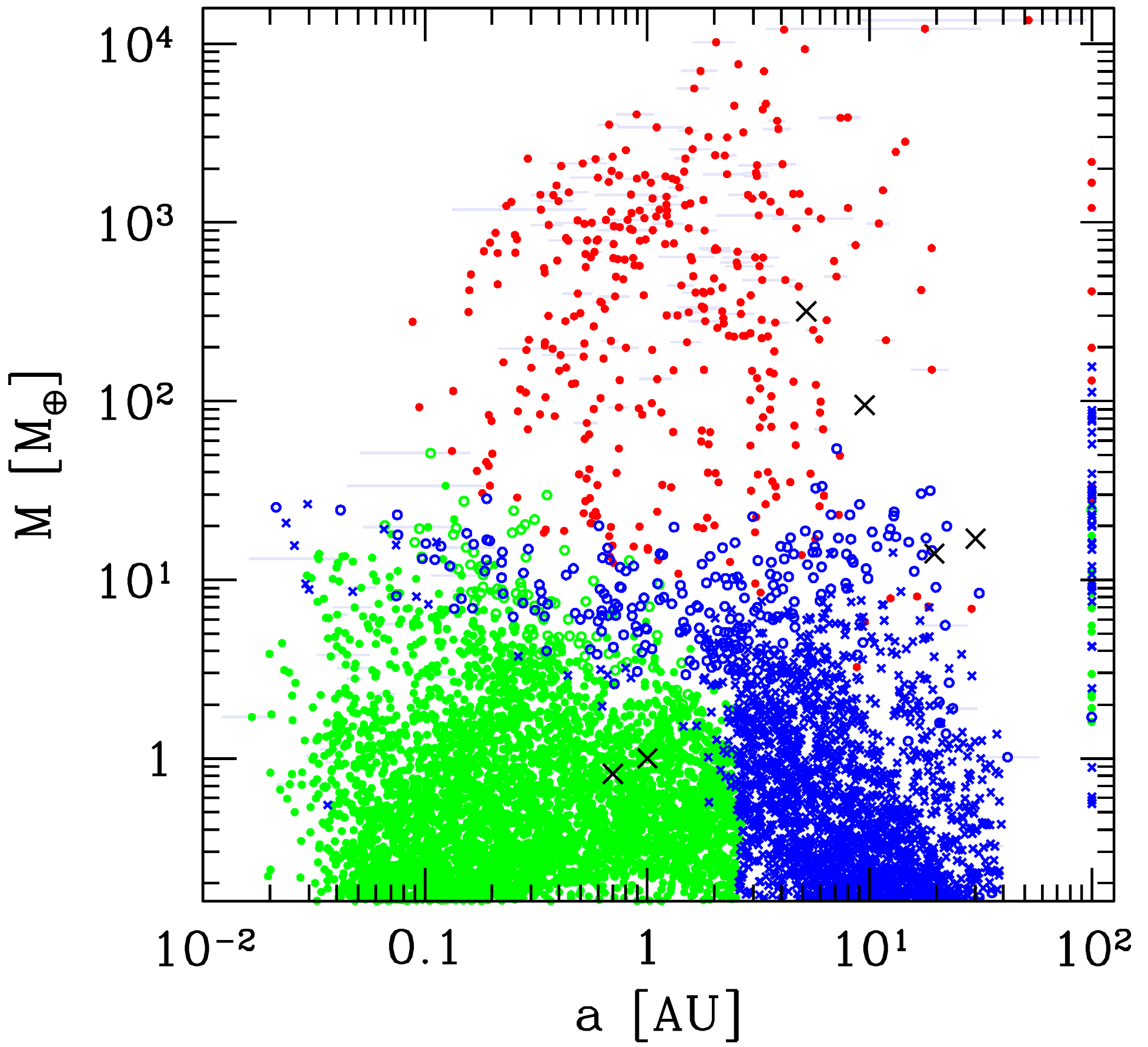}
   \caption{Synthetic mass-distance diagram. Points shows the semimajor axis, while gray horizontal bar go from $a-e$ to $a+e$. Ejected planets are shown at 100 AU. As in Fig. \ref{mamany}, the colors and symbols show the planets' bulk composition. The black crosses represent the solar system planets.}\label{ame}
\end{figure}

To finally get a statistical overview of the diversity of planetary systems, Figure \ref{ame}  shows the superposition of all 504 individual systems  in the mass-distance plane. This diagram is of similar importance for (extrasolar) planets as the Hertzsprung-Russell diagram for stars.

The first, and most fundamental results is that the variation of the initial conditions over a range indicated by observations {of protoplanetary disks} leads to a high diversity of planetary systems which covers a large part (but not all) of the parameter space that was found to be covered by the observation of extrasolar planets (compare with Fig. \ref{CMaMEpoch2017}). {The plot also visually highlights the prevalence of low-mass planets that was already a key result of the first population syntheses \citep{idalin2004a}. The prevalence is further quantified with the planetary mass function (Sect. \nameref{sect:planetarymassfunctiondistarl}) and the planet frequencies discussed in Sect. \nameref{sect:freqsynth}.} 

Considering the solar system, we see that the terrestrial planets, Jupiter, and Uranus are in regions that are well populated with synthetic planets, whereas Saturn and Neptune are rather on the outer edge of the populated envelope, at least for the 504 synthetic system shown here. As mentioned, this could be linked to a more compact early configuration of the system as predicted by the Nice Model, where Saturn and Neptune were at about 8 and 14 AU, respectively \citep{gomeslevison2005}. 

For the giant planets, a certain pile-up is see around 1 AU, similar as in the observed distribution (Sect. \nameref{sect:distribobsprop}). It is a consequence of the existence of a preferred formation location for giant planets outside of the water iceline, and a typical inward migration of  {several (1-10)} AU. {The finite extent of migration is due to the non-negligible type II migration timescale ($\approx$ viscous timescale), the slowing down in the inner system (Eq. \ref{eq:dotaown}), and the typical finite disk lifetimes that are comparable to the formation timescale of the giant planets. This mean that giants often migrate in evolved disk with a mass that has already significantly decreased, slowing down orbital migration} \citep{mordasinialibert2012a}. 

In terms of the bulk composition of the core, we see that close-in low-mass planets have an Earth-like composition (green symbols in Fig. \ref{ame}) as they did not accrete outside of the iceline. But there are also close-in, more massive (sub-)Neptunian planets that have started to form outside of the iceline giving them significant ice mass fractions indicated by blue symbols. They then migrated in through a ``horizontal branch'' \citep{mordasinialibert2009a},  as the positive corrotation torque saturates at such masses, leaving only the negative Lindblad torques, which drives fast inward Type I migration. Starting with about 50\% ice in mass in the core while accreting outside of the water iceline, they eventually have an ice mass fraction in the core of about 10-20\%, as they accrete rocky planetesimals during their migration through  the inner system, such that composition-wise, they are not really Neptune-like. This phenomenon was  previously seen in simulations for GJ 436 b \citep{figueirapont2009a}. We also see that planets with masses of about $10-30 \mearth$ have a H/He mass fraction of $\menv/\mcore$=0.1-1, and more massive planets are giants where $\menv/\mcore>1$.

\subsubsection{A quite populated planetary desert}\label{sect:populateddesert}
Compared to early population syntheses in particular from the Ida \& Lin model, there is no strong ``planetary desert'' {(absence of intermediate mass planets) visible in Fig. \ref{ame}}, even though a certain dip in the mass function at intermediate masses of about 30-100 $\mearth$ can still be seen {in the mass function} (Fig. \ref{fig:histossynth}). As partially discussed before (Sect. \nameref{sect:accretiongas}), the reason is mainly three-fold: 

First, in the models shown here, the heating from planetesimal accretion leads to a $\tKH$ that decreases less rapidly with increasing mass compared to the \citet{idalin2004a} model, as discussed in Sect. \nameref{sect:accretiongas}, meaning that planets move less rapidly through the intermediate mass regime. Thus, the probability that the gas disk disappears during this time is higher.

 Second, the gas accretion rates obtained in the disk-limited phase calculated with Eq. \ref{eq:mdotbondi} is often only a few $10^{-4}$ $\mearth$/yr. The reason is that cores often only reach a mass sufficient to trigger gas runaway in advanced stages of disk evolution, when the gas surface density has already decreased significantly. Such a timing at first appears unlikely; it is not, if we consider that in most disks, planetary growth is so slow that cores sufficiently massive to trigger gas runaway never form during the gas disk's lifetime {(the frequency of stars with giant planets is at most 20\%)}. So a late formation is actually probable. This {seems to be} a difference to pebble-based models \citep{bitschLambrechts2015}. {The removal of disks before cores have a chance to undergo runaway growth also gives naturally rise to a population of numerous super-Earths \citep{hasegawapudritz2012,alessipudritz2017}.}

Third, in contrast to earlier simulations, we find a significant multiplicity of giant planets (see Sect. \nameref{sect:freqsynth}), meaning that individual  proto-giants compete for gas while growing, reducing further the maximal gas accretion rates, and leading to more intermediate mass planets.

\subsection{The $a-R$ distribution}\label{sect:ardist}
 \begin{figure}[h]
    \centering
    \includegraphics[width=\textwidth]{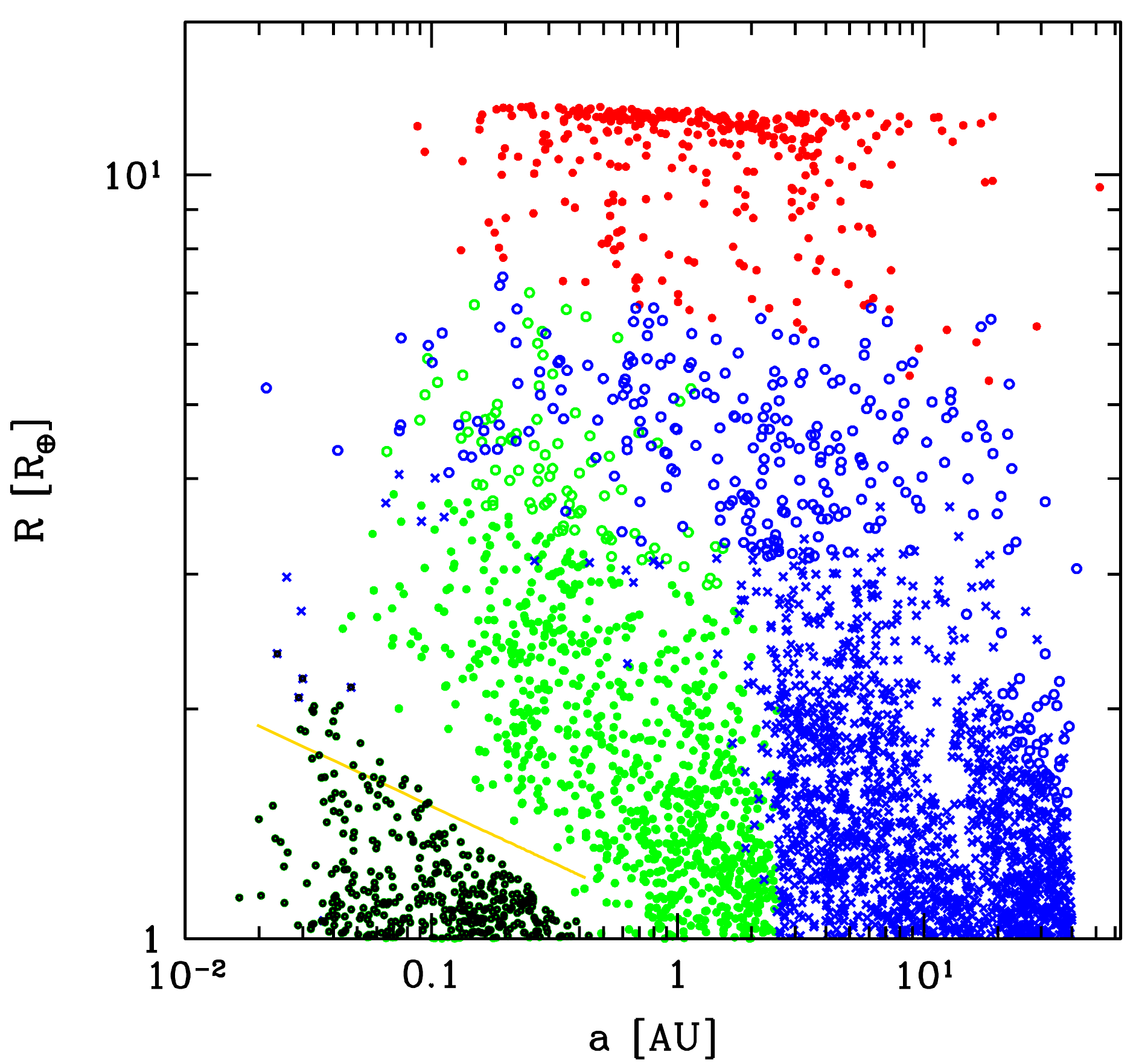}
   \caption{Synthetic planet radius-semimajor axis diagram at 5 Gyr. The colors and symbols are the same as in Fig. \ref{mamany}. Small black open circles additionally show planets that have lost the entire primordial H/He envelope by atmospheric escape. The yellow  line shows the location of the  gap determined  observationally by \citet{vaneylen2017}. }\label{ra}
\end{figure}

The results of the Kepler mission \citep{boruckikoch2011c} has given us a unique insight into the statistics of close, mostly small planets \citep[e.g.,][]{howardmarcy2012,fressintorres2013a,petigurahoward2013,mulderspascucci2015,petiguramarcy2018}. As a transit mission, it however yields planetary radii, and not masses, and no unique relation exists that links mass and radius \citep[e.g.,][]{wolfgangrogers2016}. Keeping track of the basic material type that a planet accretes (iron, silicates, ices and H/He) combined with the calculation of the evolution of its internal structure (Eq. \ref{eq:internalstruct}) make it possible to predict radii from a global model for a direct comparison \citep{mordasinialibert2012c}. 

Fig. \ref{ra} shows the synthetic distance-radius diagram at 5 Gyr.  In contrast to the mass-distance relation, there is a still signifiant evolution of the radii also after the dissipation of the gas disk because of contraction and atmospheric escape. The plot shows that giant planets with $\menv/\mcore>1$ have radii larger than 6-7 $\rearth$, while intermediate planets with $0.1\leq \menv/\mcore \leq1$ have radii larger than 3-4 $\rearth$.  To zero order, the planets of intermediate radii have a frequency that is uniform in $\log(R)$. Two prominent feature can be seen:

First, the radii of most giant planets fall in a relatively narrow range of about 10-12.4 $\rearth$ (0.9 to 1.1 $\rj${, where $\rj$ is the radius of Jupiter}). This is expected \citep{mordasinialibert2012c}, as the mass-radius relation of giant planets between about the mass of Saturn and well into the brown dwarf regime is such that the radius is nearly independent of mass, and always around 1 $\rj$  \citep{chabrierbaraffe2009b}. The reason is that the interiors become more and more compressible with increasing mass. In the synthetic population here the pile-up is  exaggerated as all planets have during evolution the same solar-composition opacity in the atmosphere \citep{freedmanlustig2014}. Varying opacities would  cause the radii to vary more \citep{burrowshubeny2007}. Additionally, no bloating effects  are included \citep[for an overview, see][]{baruteaubai2016}.

A second prominent feature is the gap running diagonally downwards with increasing semimajor axis at small radii. It separates inside of 1 AU planets that have kept or lost the primordial H/He. {Close-in low-mass planets lose their envelope because the binding energy of their H/He envelope is small compared to the incoming stellar XUV radiation that the planets are exposed to.} Note that this evaporation valley was theoretically predicted by models \citep[][]{owenwu2013,lopezfortney2013} including population synthesis models \citep{jinmordasini2014} before it was observed \citep{fultonpetigura2017,vaneylen2017}.
As expected for a simple energy-limited evaporation model with a constant efficiency factor, the model here predicts a slope that is somewhat steeper than observed \citep{owenwu2017,vaneylen2017}. The Earth-like rocky composition of the planets in the region of the gap predicted in the synthesis is consistent with the observed location of the gap \citep{owenwu2017,jinmordasini2018}.

\subsection{The planetary mass function and the distributions of $a$, $R$ and $L$}\label{sect:planetarymassfunctiondistarl}
\begin{figure}[h]
    \centering
        \begin{minipage}{0.51\textwidth}
        \includegraphics[width=\textwidth]{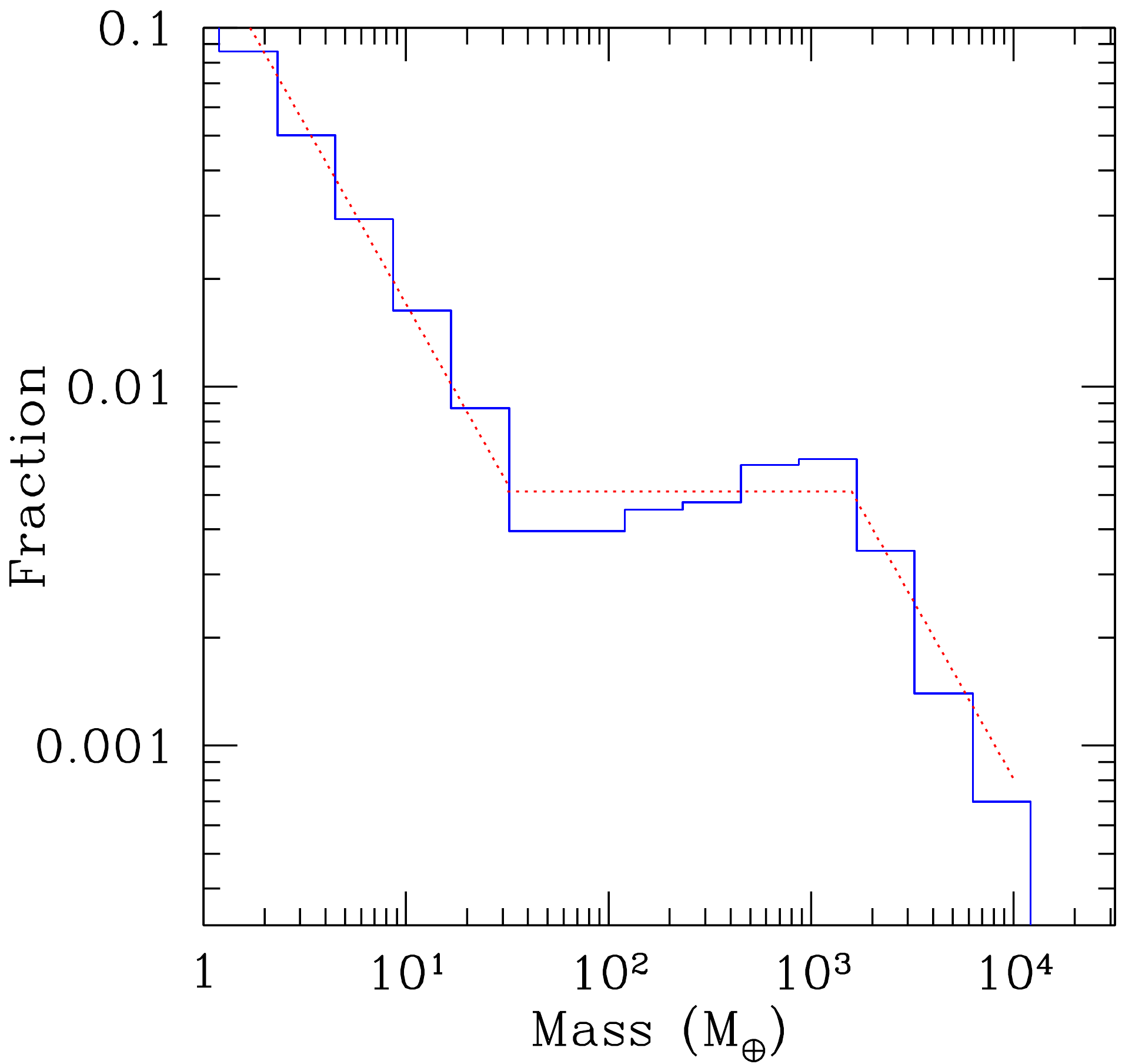}
	  \end{minipage}\hfill
        \begin{minipage}{0.49\textwidth}
        \includegraphics[width=\textwidth]{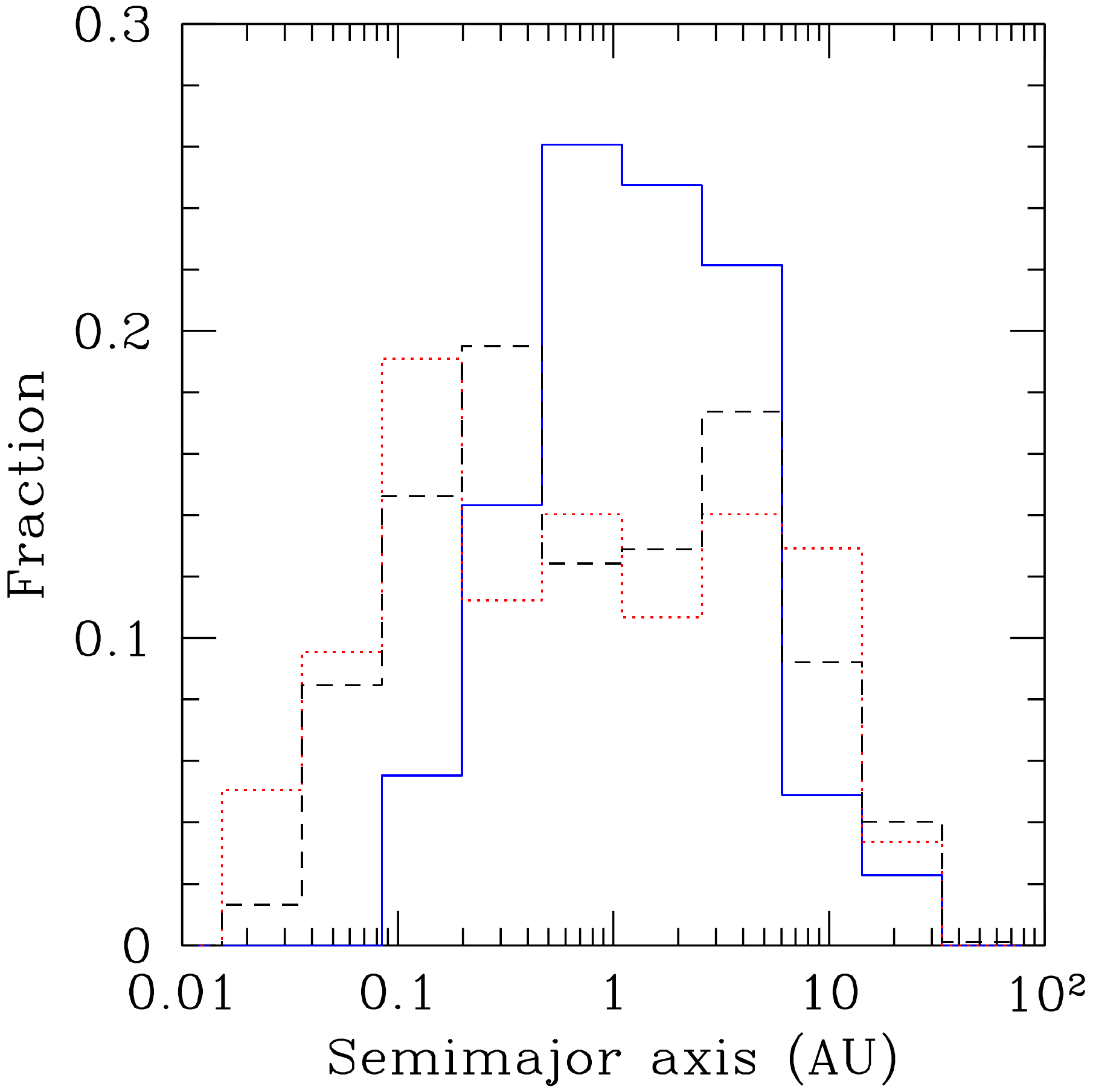}
         \end{minipage}\hfill
        \begin{minipage}{0.51\textwidth}
       \includegraphics[width=\textwidth]{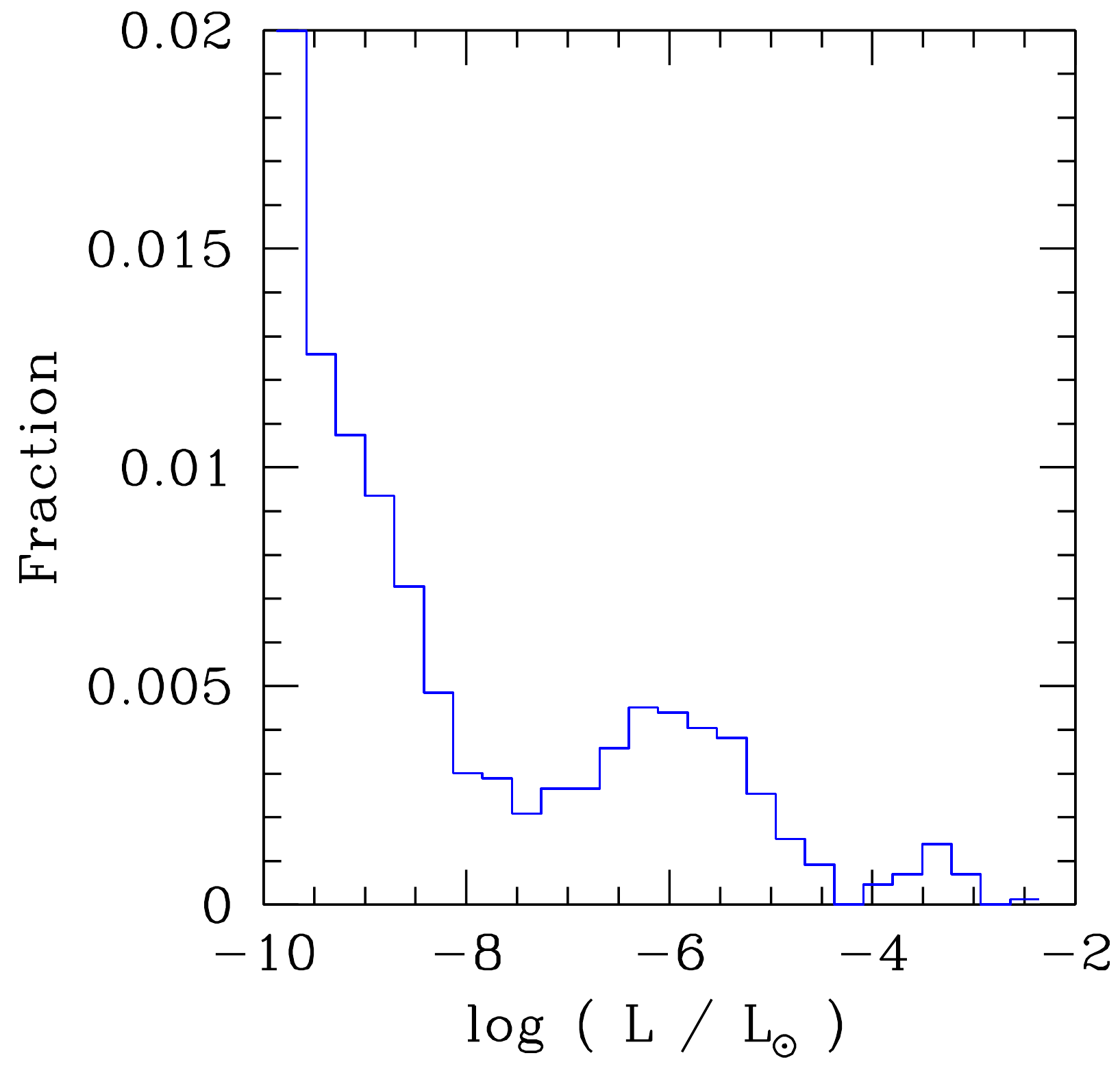}
         \end{minipage}\hfill
       \begin{minipage}{0.49\textwidth}
    \includegraphics[width=\textwidth]{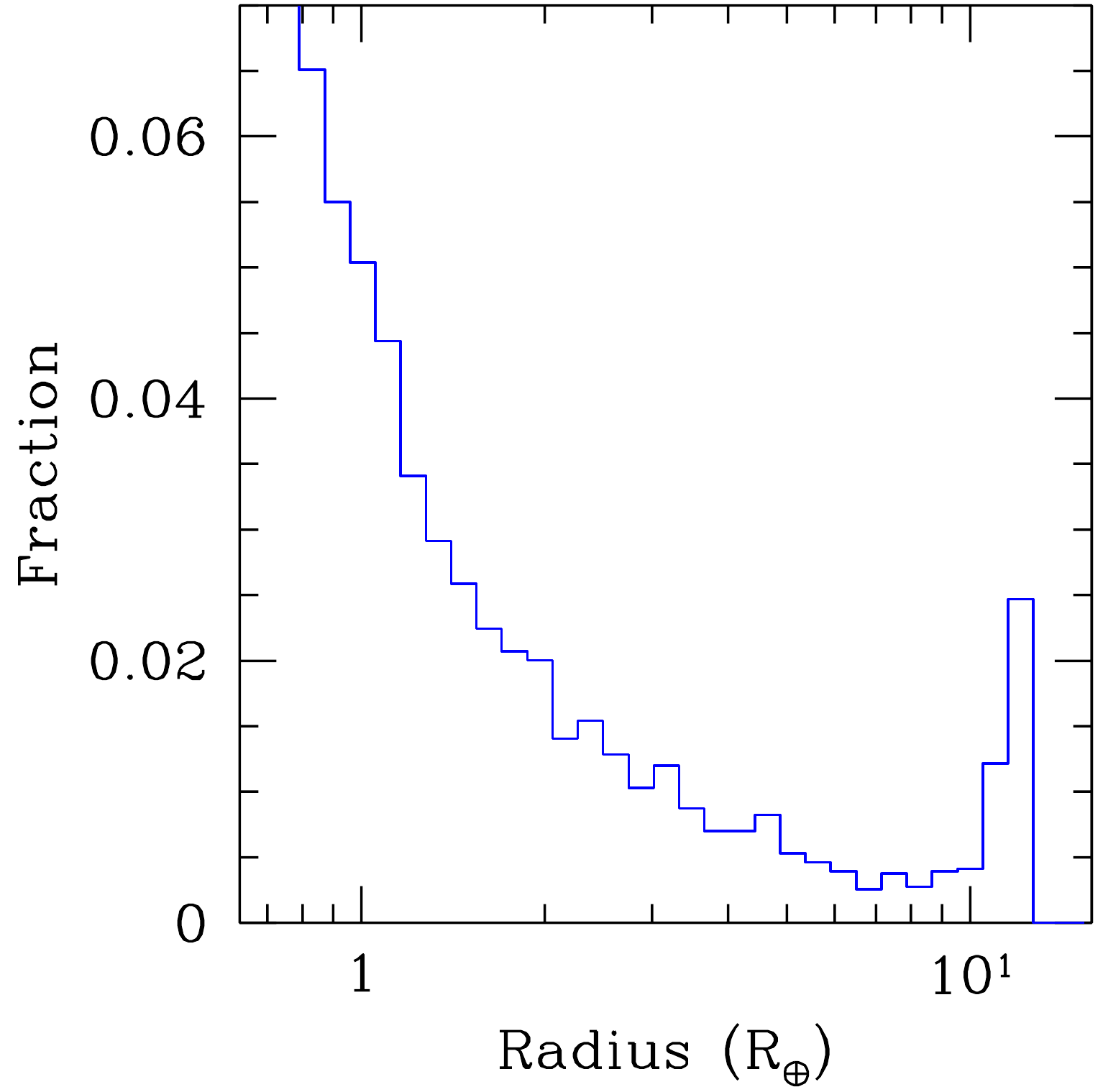}
         \end{minipage}\hfill

   \caption{Distributions of fundamental planetary properties {in the synthetic population}. Top left: planetary mass function P-MF. The red dotted lines show scalings discussed in the text. Top right: semimajor axis distribution for three mass intervals (blue $>30 \mearth$; red 10-30 $\mearth$; black 1-10 $\mearth$). Bottom left: bolometric luminosity. Bottom right: radius. The luminosity is shown at 20 Myr, the other panels are for 5 Gyr.}\label{fig:histossynth}
\end{figure}

Figure \ref{fig:histossynth} shows four  distributions of fundamental planetary properties. The top left panel shows the mass distribution (including {planets at} all semimajor axes) P-MF which is also shown with a linear y-axis in Fig. \ref{fig:histoscomp}. The prediction of the mass function is a key goal of population synthesis. We see a result that is qualitatively similar to earlier results obtained in the one-embryo-per-disk simplification \citep{mordasinialibert2009b}, and characteristic for the core accretion paradigm: two main regimes exist, one below about 30 $\mearth$ which corresponds to planets with a {composition} dominated by solids, and {one consisting of} gas-dominated giant planets at higher masses. The break at around 30 $\mearth$ corresponds to a  state when {(critical)} core and envelope mass are approximately equal, just before gas runaway accretion occurs. {The most fundamental aspect of the core accretion paradigm - the existence of a critical core mass - is thus imprinted into the planetary mass function.}

In the two regimes {below and above 30 $\mearth$} respectively, different physical mechanisms {(the accretion of solids vs. the accretion of gas) control the growth of the planets}, leading to two different slopes of the mass function. {These two  basic regimes exist both in earlier syntheses employing the one-embryo-per-disk simplification and in newer ones with numerous concurrently forming protoplanets (as here), explaining why qualitatively, the mass function is similar}.

 Above 30 $\mearth$, the mass function is to zero approximation flat in $\log(M)$ up to about 5 $\mj$, i.e., the number $N$ of planets scales with mass $M$ as $N\propto M^{-1}$. Below the break  {at about 30 $\mearth$, the dependency is steeper}, and scales roughly like $N\propto M^{-2}$.  {Finally, t}owards the upper end of the {planetary} mass function above about 5 $\mj$, the decrease follows a similar {steep} scaling. These scalings are shown in Fig. \ref{fig:histossynth} with dotted red lines.

The semimajor axis distribution of intermediate and low-mass planets is characterized by a rapid rise in the frequency between 0.01 and 0.1 AU, followed by a large interval between 0.1 and almost 10 AU where the distribution is approximately flat (or slightly decreasing) in $\log(a)$. This is similar as indicate by observations \citep{petiguramarcy2018} for $a\lesssim1$ AU. It indicates that despite orbital (Type I) migration that is included in the model without artificial reduction factors, and which leads for many planets to a significant reduction of the semimajor axis  by factors of around 4-10 or even more relative to the starting position, it nevertheless preserves the initial distribution that is uniform in $\log(a)$ as well. Giant planets are restricted to smaller semimajor axis range (about 0.1 to 6 AU), with a rapid drop both inside and outside and outside this distance. The distribution peaks a bit inside of 1 AU.

The luminosity distribution - shown at 20 Myr - mainly traces the mass distribution \citep{mordasinimarleau2017} because of the power law relation between mass and luminosity {approximately given as $L\propto M^{2}$ at a fixed time}. Compared to the mass distribution, a third local maximum appears at around $\log(L/L_{\odot})\approx-3.5$ which is caused by deuterium burning planets \citep{mollieremordasini2012}.

The distribution {of the radii of the planets} (including {planets at} all semimajor axis) finally is also similar as in simulations using the one-embryo-per-disk simplification \citep{mordasinialibert2012c}, and contains a local maximum at around 1 Jovian radius (for reasons discussed in Sect. \nameref{sect:ardist}), and a relatively continuous raise towards smaller radii. This is caused by the prevalence of low-mass planets and the fact that their KH-timescale for gas accretion is long, such that they have small H/He mass fractions and thus also small radii \citep{mordasinialibert2012c}. {This KH-timescale effect, and the EOS linking mass, bulk composition, and radius is the same as in the one-embryo-per-disk case, explaining the similarity.}
 
{Compared to the input distributions of the starting planetary embryos (initial mass of 0.1 $\mearth$ for all seeds, semimajor axes uniformly distributed in  $\log(a)$ between 0.05 and 40 AU, all seeds put into the disk at the beginning of the simulation),  the final mass and radius distributions are very different, and contain some specific physically explainable sub-structures. This suggest that the specific initial mass of the embryos  has not directly influenced these distributions, at least for planets with $M\gg0.1\mearth$. On the other hand, the final semimajor axis distribution is still -to zero order- uniform in $\log(a)$ for  planets with masses between 1 and 30 $\mearth$. This could indicate that the input semimajor axis distribution influences this result, and that the final distribution would differ if another initial distribution would be used. Such different distributions seem quite possible, for example because of preferred formation locations of the planetesimals \citep[e.g.,][]{drazkowskaalibert2016}, particle pile-ups outside of orbits of already existing planets \citep[e.g.,][]{pinillabirnstiel2015} or strong migration traps \citep[e.g.,][]{hornlyra2012,hasegawapudritz2012}. This shows the necessity to include the earlier stages of planet formation in future global models, namely the dust, pebble and planetesimal formation stages.} 

\subsection{Comparison with observations: planet frequencies}\label{sect:freqsynth}
Before comparing observed and synthetic distributions, we address the frequency of three fundamental planet types predicted by the synthesis: first, giant planets with a mass of at least 300 $\mearth$, second close-in planets (period $\leq100$ d, i.e., $a\leq0.42$ AU, radius $\geq1\rearth$) comparable to the planets probed by Kepler \citep[e.g.,][]{marcyweiss2014,petiguramarcy2018}, and third planets in the classical habitable zone with mass of  0.3 to 5 $\mearth$ and a semimajor axis of 0.95 to 1.37 AU \citep{kastingwhitmire1993}. We give the overall fraction of stars having such planets, indicating also their multiplicity. For the comparison with observed frequencies (Sect. \nameref{sect:frequenciesobs}), one should  keep in mind that the  {synthetically predicted} absolute frequencies are less robust than the relative {ones. The reason is that the absolute frequencies depend more directly on model parameters like the arbitrary chosen planetesimal size of 300 m which influences the solid accretion rate, as discussed in Sect. \nameref{sect:accretionofsolids}. Assuming  a smaller (bigger) size of the planetesimals would increase (decrease)  the fraction of disks in which massive planets form. } 

\begin{table}
\caption{Percentage of stars with N planets of the given type {in the synthetic population, and comparison to observations (last line)}.}
\label{tab:frequenciessynth}   
\begin{tabular}{p{2.4cm}p{2.cm}p{2.cm}p{2.cm}}
\hline\noalign{\smallskip}
 N& Giant planets & Close-in planets & Planets in HZ   \\
\noalign{\smallskip}\svhline\noalign{\smallskip}
1            & 4.8 & 8.4 & 30.7  \\
2           & 7.4 & 12.8&   8.2\\
3           &  5.4& 11.4  & 1.0  \\
4           & 0.4 & 10.0 & 0.0 \\
$\geq$5     & 0.0 & 11.4 & 0.0  \\
Overall synthetic & 18.0 & 54.0 & 39.9  \\ \hline
{Overall observed} & {10-20}& {50-60}&  {unknown}\\
\noalign{\smallskip}\hline\noalign{\smallskip}
\end{tabular}
\end{table}
We see that the overall fraction of stars with giant planets is about 18\%. The most frequent number of giant planets per star is 2, occurring for 7.4\% of all stars. Single giant planets and stars with 3 giants are both on the 5\% level. Only 0.4 \% of the stars have 4 giant planets, and none has more than that. Note that these numbers may be upper limits, as the calculation of the N-body interaction was stopped at 10 Myr, such that we do not take into account ejections or collisions at later moments. The overall frequency of giant planets is however similar as observed ({10-20\%, see} Sect. \nameref{sect:frequenciesobs}), even if the multiplicity in the synthetic population might be higher than observed \citep{bryanknutson2016}.  

The population of close-in planets is dominated by small (or low-mass) planets as visible from Figs. \ref{ame} and \ref{ra}. Even if a quantitative comparison with the HARPS high-precision RV survey or the Kepler transit survey would require a dedicated modeling of the observational biases \citep[e.g.,][]{mayormarmier2011,petiguramarcy2018}, the frequencies of these close-in planets in Table \ref{tab:frequenciessynth} makes nevertheless clear that such planets are a very common outcome of the formation process, similarly as observed (Sect. \nameref{sect:frequenciesobs}). About 54\% of the synthetic systems have such planets {which is similar to the observed frequency \citep{petigurahoward2013}}, and the multiplicity is high, with a mean number of about 3 such planets per star that has this type of planet, which is again at least qualitatively similar as observed. This is high frequency of close-in planets can only be reproduced in the syntheses if a steep, centrally concentrated distribution of the planetesimals is used, as described in Sect. \nameref{sect:initialconditionssynthesis} (see also \citealt{chianglaughlin2013}). This is an interesting constraint for drift and planetesimal formation models \citep[e.g.,][]{drazkowskaalibert2016}.

The fraction of stars with planets in the classical habitable zone is in contrast lower, but with 39.9\% still very significant. The mean number of this type of planet is 1.25 for those stars that have such planets, i.e. typically there is only one planet in the classical habitable zone per such system. {Observationally, the frequency of solar-like stars with potentially habitable planets is still now well known, with estimates ranging from 1 to 100\% \citep[e.g.,][]{burkechristiansen2015}. For M-dwarfs, where a direct determination of this frequency is in contrast already possible with radial velocity surveys, the fraction is $0.41^{+0.54}_{-0.13}$ \citep{bonfilsdelfosse2013}. } 

\subsection{Comparison with observations: distributions}\label{sect:compobsdist}
\begin{figure}[h]
    \centering
        \begin{minipage}{0.51\textwidth}
        \includegraphics[width=\textwidth]{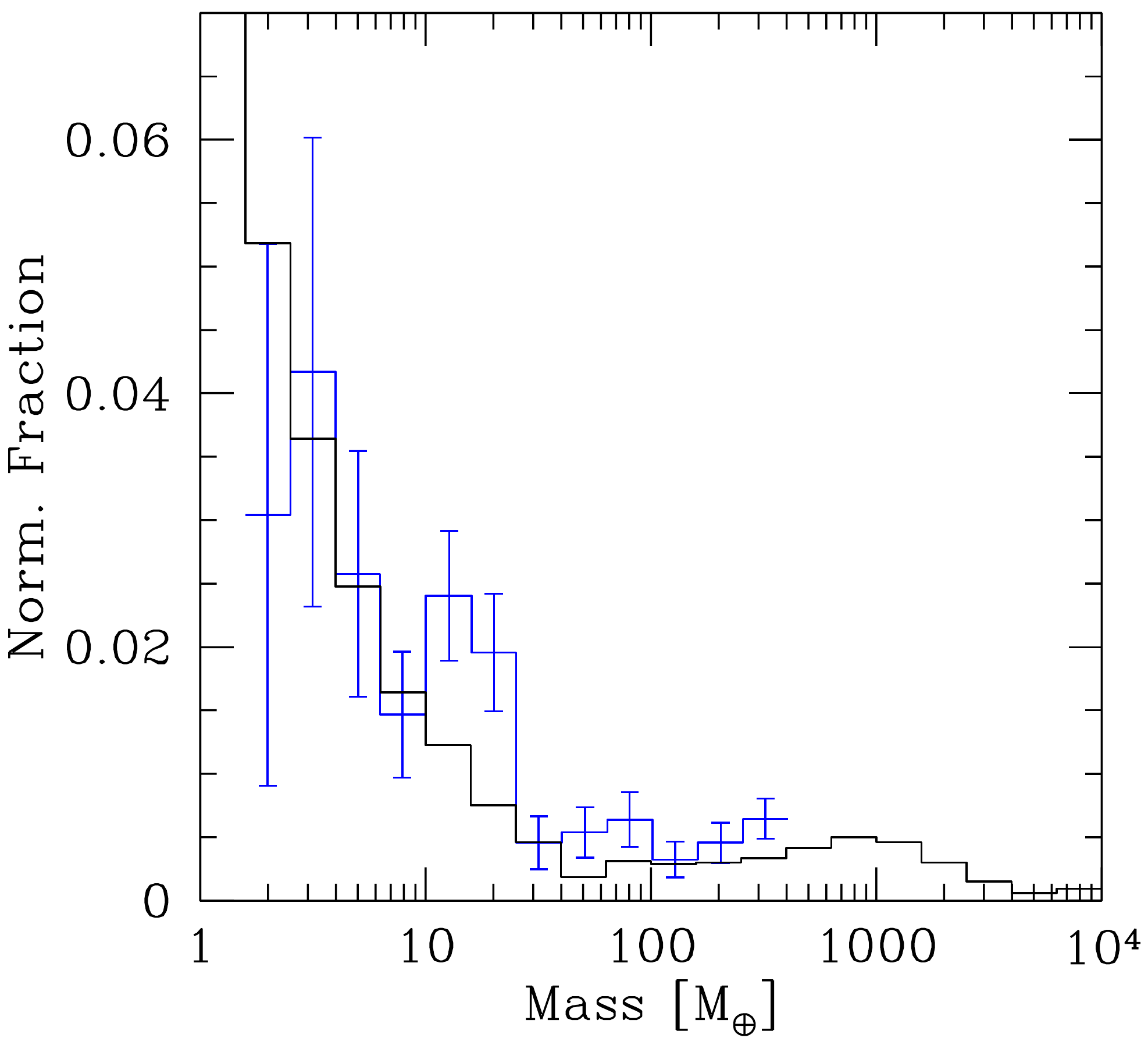}
	  \end{minipage}\hfill
        \begin{minipage}{0.49\textwidth}
        \includegraphics[width=\textwidth]{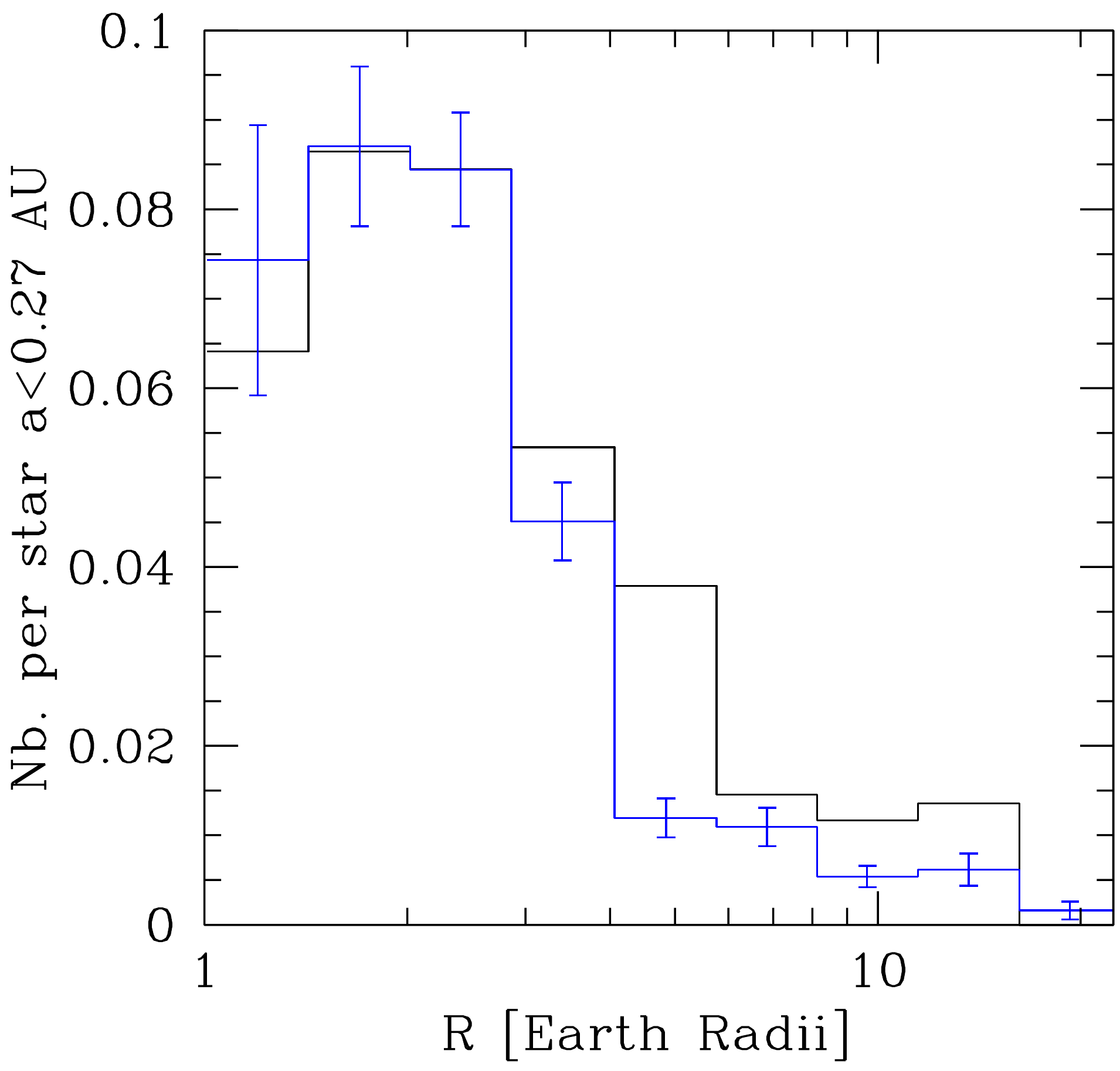}
         \end{minipage}\hfill

   \caption{Comparison of the synthetic mass and radius distribution  (black) {of the population introduced at the beginning of Sect. \nameref{sect:results}} with the bias-corrected observed distributions (blue) of the HARPS high precision RV survey \citep{mayormarmier2011} and the Kepler transit survey \citep{howardmarcy2012}. The two theoretical and observed distributions were normalized to the same value at the bin at 30 $\mearth$ and   2-3 $\rearth$, respectively.}\label{fig:histoscomp}
\end{figure}

Figure \ref{fig:histoscomp} compares the synthetic mass and radius distribution with their observational counterparts as found by the HARPS high precision RV survey \citep{mayormarmier2011} and an analysis of the Kepler transit survey \citep{howardmarcy2012}. The observed distributions are corrected for the observational bias. {The synthetic distributions only include the planets in the same mass/radius and orbital distance range as in the observational samples.}

The basic shape echoes the distributions {that include all synthetic planets shown} in Fig. \ref{fig:histossynth}. The {synthetic mass} distribution  compares quite well with the observed one, in particular regarding the aforementioned break in the mass function at about 30 $\mearth$ and the associated change of the slope that was already predicted in early population syntheses \citep{mordasinialibert2009b}. This break is visible also in other bias-corrected high-precision RV surveys like \citet{howardmarcy2010}, and even in the biased, directly observed mass distribution  \citep{schneiderdedieu2011,wrightfakhouri2011}. The log-flat distribution between about 30 $\mearth$ and 5 $\mj$ agrees with the observed distribution \citep{marcybutler2005} {as well}. The fact that the two theoretically predicted slopes are  visible also in the observed distribution, and that the break occurs at a similar mass in theory and observations constitutes a major success of the core accretion theory{, and planet formation theory in general. This is of an astrophysical importance comparable to the development of  a theory for the stellar initial mass function \citep{chabrier2003}, including the classical Salpeter slope  \citep{salpeter1955}}. 

In the  radius distribution of synthetic planets inside of 0.27 AU,  a significant difference is seen for the peak at about 1 Jovian radius relative to the distribution that includes all orbital distance in Fig. \ref{fig:histossynth}, as already found in \citet{mordasinialibert2012c}: the peak is less pronounced compared to Fig. \ref{fig:histossynth} because of the broader bins, and due to the fact that only planets inside of 0.27 AU are included (as in the observational sample), while synthetic giant planets are mostly further out (see Fig. \ref{ra}). In these broad bins, the evaporation valley and the associated gap in the radius distribution is not visible as a finer radius resolution is required \citep{jinmordasini2018}. Another difference is that the synthetic distribution is less abruptly increasing towards the small radii compared to the observed one.

We note that the population contains many planets with masses still lower than 1 $\mearth$ and/or radii less than 1 $\rearth$. They {are found in} systems in which not much growth and migration occurred during the first 10 Myr {(the time during which the systems' formation was simulated as explained in Sect. \nameref{sect:initialconditionssynthesis})}. At {10 Myr}, there are still around 20 low-mass protoplanets left. {Such systems arise for initial conditions with low amounts of solids (low [Fe/H] and/or initial gas disk masses) and/or short disk lifetimes (see \citealt{mordasinialibert2012a} for an extensive discussion of the correlations between disk and planetary properties). This dependency is illustrated by the systems forming at low [Fe/H] in Fig. \ref{mamany}. }

Over longer timescales, {these low-mass planets} could collide to form more massive planets, such that this result could be an artifact of only modeling the N-body interaction during 10 Myr. This also means that all results concerning planets with masses close to the initial embryo mass should be taken with caution.

\subsection{Correlations with disk properties}\label{corrdiskprops}
An important application of population synthesis is to understand how the planetary formation process depends on the properties of the protoplanetary disk. The most important observed correlation in this context is that the probability of observing giant planets increases with the host star metallicity \citep[e.g.,][]{gonzalez1997,santosisraelian2004,fischervalenti2005}, the so-called ``metallicity effect'', as discussed in Sect \nameref{sect:obscorrstellarprops}. Syntheses by \cite{idalin2004} and later \citet{mordasinialibert2009b,mordasinialibert2012a} have shown quantitatively that this metallicity effect is a natural outcome of the core accretion model. This is under the assumption that stellar and disk metallicity are proportional (Sect. \nameref{sect:initialconditionssynthesis}), and further, that the surface density of planetesimals increases with the mass fraction of heavy elements as well (Eq. \ref{eqSigmad}), which is indicated by planetesimal formation models \citep[][]{brauerdullemond2008}. In this case, the growth of a planetary core by accreting planetesimals occurs on a shorter timescale (Eq. \ref{eq:taucoreaccr}), and leads to more massive cores as well \citep{kokuboida2012}. Thus, there is a higher chance to grow to the critical core mass and trigger rapid gas accretion before the gas disk has dissipated.

 \begin{figure}[h]
    \centering
    \includegraphics[width=0.5\textwidth]{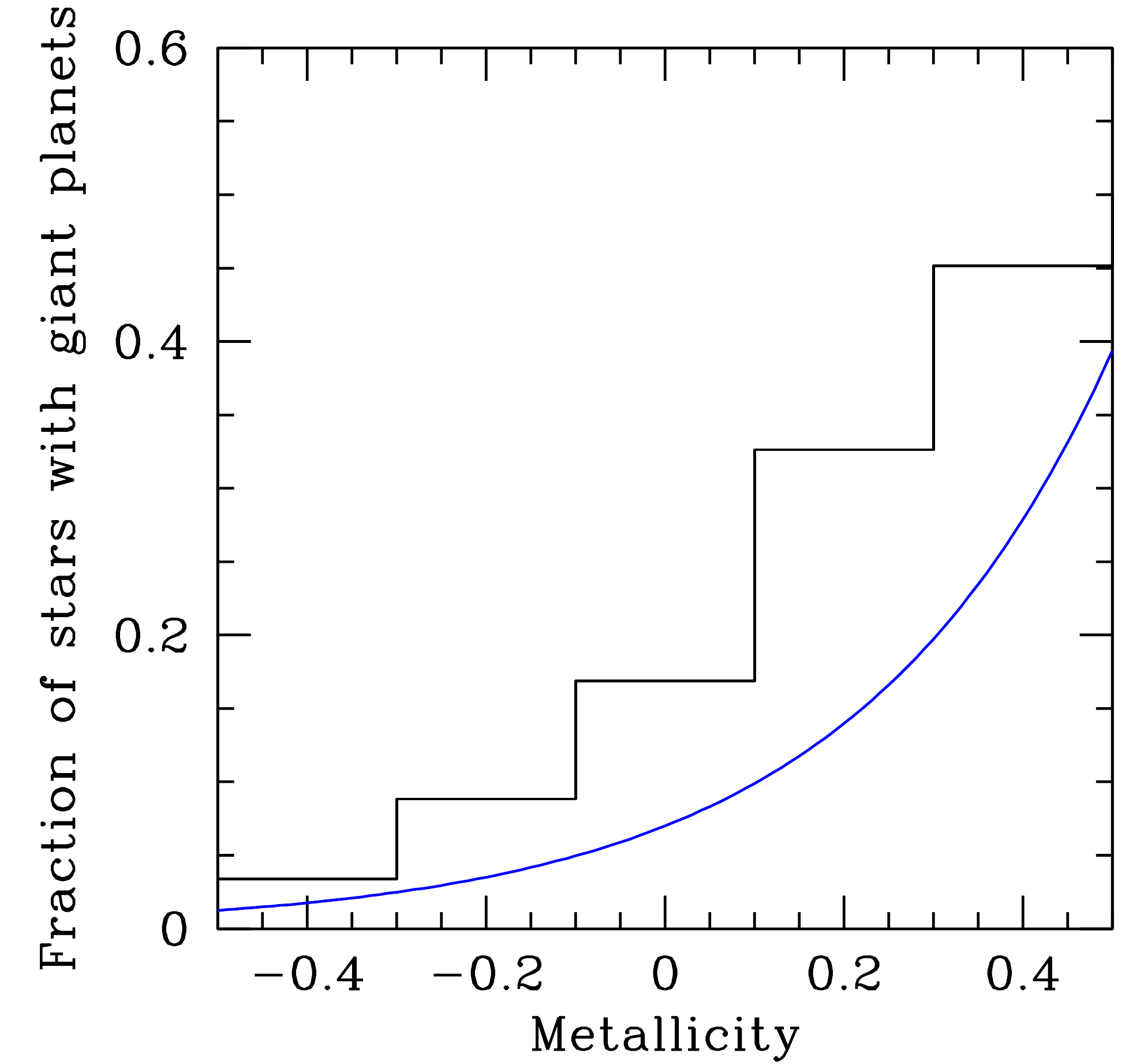}
   \caption{Fraction of stars with giant planets ($M\geq300\mearth$) as found in the synthetic population (black line). The blue line shows the fit to the observed frequency from \citet{mortiersantos2013}.}\label{metallicityeffect}
\end{figure}

Figure \ref{metallicityeffect} shows the fraction of stars with at least one giant planet (mass higher then 300 $\mearth$) in the present synthesis for 1 $\msun$ stars as a function of metallicity, compared to a fit to the observed relation by \citet{mortiersantos2013}. While the model over-predicts the  number of giant planets in absolute terms, it agrees with the observed relative increase quite well.

Further correlation of disk properties were studied by \citet{mordasinialibert2012a} where - not surprisingly - a high number of correlations was found: for example, for the planetary initial mass function, high metallicities lead to a higher frequency of giant planets while higher initial disk (gas) masses lead mainly to giant planets of a higher mass. For long disk lifetimes, giant planets are both more frequent and massive. At low metallicities, very massive giant planets cannot form, but otherwise giant planet mass and metallicity are nearly uncorrelated. In contrast, (maximum) planet masses and disk gas masses are correlated.

\subsection{Testing theoretical sub-models}\label{sect:testingsubmodels}
The final goal of population synthesis is to improve our understanding of planet formation and evolution. For this task, specific sub-models are put via syntheses to the observational test as described in Sect. \nameref{sect:workflow}.  A non-conclusive list of mechanisms that were addressed in this way is (1) orbital migration, mostly (non)isothermal Type I migration. {Early population syntheses \citep{idalin2008c,mordasinialibert2009b} showed that the observed distribution of planetary orbital distances and the fraction of stars with hot Jupiters can only be reproduced if the migration rates predicted by the then existing (isothermal) type I migration models \citep{tanakatakeuchi2002} are strongly reduced. This sparked numerous dedicated studies that led to more realistic non-isothermal type I migration rates \citep[like ][]{massetcasoli2010,kleybitsch2009,paardekooperbaruteau2010}. These new models where then in turn included in the population syntheses, leading to orbital distances that are more similar to observations  \citep{dittkristmordasini2014}. This is a prime example of how population synthesis and specialized models  advance each other.} (2) grain dynamics and opacities in protoplanetary atmospheres influencing the bulk composition of planets. {Here it was found with population syntheses that the observed mass-radius relation (i.e., the bulk composition) of extrasolar planets with H/He can be reproduced only if the grain opacity in  protoplanetary atmospheres is clearly lower than the ISM opacity \citep{mordasiniklahr2014}. This led to the development of specialized models for the grain dynamics (growth, settling, destruction) and resulting opacities \citep{ormel2014,mordasini2014}, which are indeed low compared to the ISM.}  (3) disk inhomogeneities and transitions leading to migration traps \citep{hasegawapudritz2011a,hasegawapudritz2013,colemannelson2016}, (4) stellar cluster environments \citep{ndugubitsch2018}, (5) the gas accretion shock structure  and the luminosity of young giant planets  \citep{mordasinimarleau2017}, (6) constraints on  formation  pathways from the chemical composition and atmospheric spectra \citep[e.g.,][]{marboeufthiabaud2014,madhusudhanamin2014}, or finally (7) atmospheric escape of primordial H/He envelopes \citep{jinmordasini2014,jinmordasini2018}.

\subsection{Predictions{: observational confirmations and rejections}}\label{sect:predictionsconfirmationrejection}
Another central application of population syntheses are predictions for upcoming instruments and surveys, {i.e., quantitative theoretical predictions that can be falsified with more accurate observational methods}. \citet{mordasinialibert2009b} for example studied the consequences for the fraction of stars with detectable planets and the shape of the P-MF if the radial velocity measurement accuracy improves from 10 m/s to 1 m/s {and 0.1 m/s.}

{Some of the predictions made by population syntheses were later confirmed by observations, others turned out to be inconsistent. Some of  the most important confirmed predictions are:} (1) the prevalence of low-mass and small planets \citep{idalin2004} {that was made well} before they were found by precise RV surveys and Kepler \citep[e.g.,][]{howardmarcy2010,mayormarmier2011,boruckikoch2011c}, (2) the break in the planetary mass function at around 30 $\mearth$ \citep{mordasinialibert2009b} {that was} later detected by \citet{howardmarcy2010,mayormarmier2011} through high-precision RV, (3) the pile-up of planetary radii around 1 $\rj$ \citep{mordasinialibert2012c} which was not visible in early polluted  Kepler data but which is has become apparent recently in cleaned samples \citep{petiguramarcy2018}, or (4) together with other models \citep{owenwu2013,lopezfortney2013}, the depleted evaporation valley in the distance-radius plane \citep{jinmordasini2014}, which was later {confirmed observationally} by \citet{fultonpetigura2017} {and \citet{vaneylen2017}}. 

{Important predictions that turned out to be inconsistent with observations were (1) the existence of a strongly depleted ``planetary desert'' \citep{idalin2004}, i.e., a strong depletion in the frequency of planets with masses between 10 to 100 $\mearth$ (a weak depletion might  actually exist, see Fig. \ref{CMaMEpoch2017}), or (2) an absence of close-in planets ($a\lesssim0.1$ AU) with masses less than approximately 10 $\mearth$ \citep{mordasinialibert2009a}. These inconsistent predictions are actually of particular interest, as they point at important shortcomings in the theoretical models. In the former case, the gas accretion rate in the runaway phase was overestimated (see Sect. \nameref{sect:accretiongas} and \nameref{sect:populateddesert} for limiting effects), in the latter, a strongly reduced isothermal type  I migration rate and a criterion for the transition into type II migration  based only on the thermal criterion caused the discrepancy. These shortcomings were then addressed in later generations of the models (Sect. \nameref{sect:overviewpopmodels}), and helped in this way to improve the understanding of the planet formation process, and to avoid oversimplifications.}

\section{{Summary and conclusions}}
{The increase in statistical observational constraints on extrasolar planets has been enormous in the last two decades. Both ground and space-based surveys have derived distributions of fundamental planetary properties like the frequency of planets in the mass-distance and radius-distance planes, the planetary mass function, the eccentricity distribution, or the planetary mass-radius relation.} 

{All these observed distributions put strong statistical constraints on the theory of planet formation and evolution (Sect. \nameref{sect:statobsconst}).  The method of choice to use these constraints in order to improve our understanding of planet formation is populations synthesis, an approach that has been used for many decades in various fields of stellar astrophysics.}

{The underlying idea of population synthesis  is that the same physical processes govern the formation of planets in all protoplanetary disk, but that the initial conditions for these processes (the properties of the parent protoplanetary disk) and potentially the boundary conditions (like the stellar cluster environment) differ, and that this gives raise to the observed diversity of (extrasolar) planets.}

{Methodically, population syntheses (Sect. \nameref{sect:popsynthmethod}) thus consist of two main components: first, probability distributions of the initial conditions (disk properties, Sect. \nameref{sect:initialconditionssynthesis}) and second a global end-to-end model of planet formation and evolution that can predict observable planetary properties based directly on disk properties.  Building both on the core accretion and the gravitational instability scenario, the different modern population synthesis models in the literature (Sect. \nameref{sect:overviewpopmodels}) include in a self-consistently coupled way an impressive number of physical processes like the evolution of the protoplanetary disks of solids and gas, planetary accretion of solids (both of planetesimals and pebbles), the accretion of gas, orbital migration, and N-body interactions, and often several more. A description of these sub-models can be found in Sect.  \nameref{sect:globalmodels}.}

{The sub-models describing these processes are, however, either parameterized or low-dimensional static approximations of 3D dynamical systems (Sect. \nameref{sect:lowdimapprox}). To what extent the dynamical multi-dimensional nature of one individual governing process (like orbital migration or pebble accretion) can be ``distilled'' into a simpler, lower-dimensional approximation that still captures the essence of the physics while allowing population syntheses with acceptable computational costs, is a key challenge for any population synthesis approach, and an ongoing development.} 

{On the other hand, population synthesis is often the only possibility to observationally test theoretical models of a specific mechanism (Sect. \nameref{sect:testingsubmodels}) as it produces synthetic data that can be directly statistically compared with observations, factoring in the non-linear interaction between the different mechanism concurrently acting during planet formation. This leads to planetary formation tracks and synthetic planetary systems of a large diversity (Sect. \nameref{sect:diversityofplanetarysystemarchitectures}). Population synthesis therefore also has a high predictive power that is difficult to achieve with theoretical models of just one physical process.}

{In the Section \nameref{sect:results}, the typical output obtained from a population synthesis model is presented and linked to underlying physical effects. These are the distribution of planets in the mass-distance and radius-distance plane, the planetary mass function, and the distribution of radii, orbital distances, and luminosities. These distributions are compared with their observed counterparts (Sect. \nameref{sect:compobsdist}).}

{Among these results, a first key prediction of population synthesis models (Sect. \nameref{sect:amdistribution}) was the prevalence of low-mass and small planets  that was later observationally confirmed by high-precision radial velocity surveys and the Kepler satellite (Sect. \nameref{sect:predictionsconfirmationrejection}). Clearly this has important implications beyond planet formation theory like for the question about the existence of other habitable planets.}

{A second key prediction was the existence of two regimes in the planetary mass function (Sect. \nameref{sect:planetarymassfunctiondistarl}) with a break at around 30 $\mearth$ when the transition from solid to gas-dominated planet occurs at the critical core mass. The most fundamental aspect of the core accretion paradigm - the existence of a critical core mass - is thus imprinted into the planetary mass function. This prediction was later confirmed by high-precision RV surveys (Sect. \nameref{sect:predictionsconfirmationrejection}). The finding that the two theoretically predicted slopes in the mass function are visible also in the observed distribution, and that the observed break occurs at a similar mass as theoretically predicted represents a major success for the core accretion theory, and of planet formation theory in general. This is in a broader astrophysical context of similar importance as the development of a theory for the stellar initial mass function. }

{A third, maybe even more fundamental insight is that the population syntheses show that the variation of the initial conditions over a range suggested by protoplanetary disk observations lead to an extreme diversity in the resulting synthetic planetary systems, which is probably the single most characteristic property also of the actual extrasolar planet population. It indicates that at least some of the strong non-linearities and feedback mechanisms occurring during planet formation are indeed captured in the theoretical models.}

{These points do, however, clearly not mean that the current population models describe in a definitive way the actual planet formation and evolution process. They rather reflect the state of the field of planet formation theory where important  physical mechanisms governing planet formation are still not well understood (Sect. \nameref{sect:confrontingtheoryobs}). It is therefore not surprising that other important predictions made by population synthesis models turned out to be inconsistent with observations (Sect. \nameref{sect:predictionsconfirmationrejection}). But it is exactly the quantitative falsifability of population synthesis that is important, as this it is the key to reject some theoretical concepts or to identify missing ones, improving in this way our understanding of  how planets form and evolve. }

\vspace{1cm}

\small{Acknowledgements: C.M. thanks Ralph Pudritz for the invitation to write this review and for the editorial guidance. C.M. acknowledges the support from the Swiss National Science Foundation under grant BSSGI0$\_$155816 ``PlanetsInTime''. Parts of this work have been carried out within the frame of the National Center for Competence in Research PlanetS supported by the SNSF.}

\bibliographystyle{aa}
\bibliography{liball.bib}{}

\end{document}